\documentclass[twocolumn,amsmath,amssymb,floatfix,superscriptaddress,10pt
,aps,pre]{revtex4-1}

\newcommand{\bvec}[1]{{\mathbf{\string#1} }}
\newcommand{\upd}{\mathrm{d}}
\usepackage{graphicx}
\usepackage{amssymb,amsfonts,amsmath}
\usepackage{color}
\usepackage{ulem}
\usepackage{fancybox}

\newcommand{\taua}{\tau}

\newcommand{\Da}{\mathcal{D}_\text{a}}

\DeclareSymbolFont{matha}{OML}{txmi}{m}{it}
\DeclareMathSymbol{\varv}{\mathord}{matha}{118}

\newcommand{\di}{\mathfrak{d}}

\newcommand{\UCNA}{\text{(u)}}
\newcommand{\FOX}{\text{(f)}}
\newcommand{\BFPA}{\text{(b)}}

\begin{document}

\title{
Confined active particles with spatially dependent Lorentz force: an odd twist to the ``best Fokker-Planck approximation''}
\author{Ren\'e Wittmann}
\email{rene.wittmann@hhu.de}
 \affiliation{Institut f\"{u}r Theoretische Physik II: Weiche Materie, Heinrich-Heine-Universit\"{a}t D\"{u}sseldorf, D-40225 D\"{u}sseldorf, Germany}
 \affiliation{Institut für Sicherheit und Qualität bei Fleisch, Max Rubner-Institut, D-95326 Kulmbach, Germany}
 \author{Iman Abdoli}
 \affiliation{Institut f\"{u}r Theoretische Physik II: Weiche Materie, Heinrich-Heine-Universit\"{a}t D\"{u}sseldorf, D-40225 D\"{u}sseldorf, Germany}
 \author{Abhinav Sharma}
\affiliation{Mathematisch-Naturwissenschaftlich-Technische Fakult\"at, Institut f\"ur Physik, Universit\"at Augsburg, Universit\"atsstra{\ss}e 1, D-86159 Augsburg, Germany}
\affiliation{Leibniz-Institut f\"ur Polymerforschung Dresden, Institut Theory der Polymere, D-01069 Dresden, Germany}
\author{Joseph M.\ Brader}
 \affiliation{Department of Physics, University of Fribourg, CH-1700 Fribourg, Switzerland}

\date{\today}

\begin{abstract}
We derive a version of the so-called ``best Fokker-Planck approximation'' (BFPA) to describe the spatial properties of interacting active Ornstein-Uhlenbeck particles (AOUPs)  in arbitrary spatial dimensions.
In doing so, we also take into account the  odd-diffusive contribution of the Lorentz force acting on a charged particle in a spatially dependent magnetic field,  sticking to the overdamped limit.
While the BFPA itself does not turn out to be widely useful, our general approach allows to deduce an appropriate generalization of the Fox approximation,
which we use to characterize the stationary behavior of a single active particle in an external potential by deriving analytic expressions for  configurational probability distributions (or effective potentials).
In agreement with computer simulations, our theory predicts that the Lorentz force reduces the effective attraction and thus the probability to find an active particle in the vicinity of a repulsive wall.
Even for an inhomogeneous magnetic field, our theoretical findings provide useful qualitative insights, specifically regarding the location of accumulation regions.
\end{abstract}

\maketitle

\section{Introduction}

For over a decade, the peculiar and diverse nonequilibrium behavior of active particles,
ranging from swimming organisms over Janus colloids to vibrated granular media 
 has stimulated an immense body of research \cite{tevrugt_REVREV}, including the development of numerous theoretical approaches~\cite{REVromanczuk2012,REVbechinger2016}.
 Specifically, ever since the first efforts~\cite{maggi2015sr,szamel2014} to describe the self-propulsion of active colloids by Ornstein-Uhlenbeck processes \cite{OrnsteinU},
such active Ornstein-Uhlenbeck particles (AOUPs) have been a particularly convenient model system for theoretical analysis~\cite{szamel2015,flenner2016nonequilibrium,szamel2017evaluating,berthier2017,flenner2020departure,bonilla2019,marconi2017_expansion,fodor2016,martin2020,martin2021_expansion,fily2017,fily2019,caprini2018linRESP,caprini2019entropyPROD,caprini2020velocityCORR1d,caprini2021FDR,dabelow2019,woillez2020letter,woillez2020,bothe2021,semeraro2021work,semeraro2023work}
as their self-propulsion is described as a Gaussian process.
 The continuously increasing popularity of  the relatively simple AOUP dynamics \cite{fily2012athermal,sandford2017,sandford2018,activeCurvature,singh2020,dalcengio2021,maggi2022_AOUPfield,caprini2022spatial1,caprini2022spatial2,keta2022disordered,keta2023intermittent,keta2024emerging,davis2024control,zheng2024quantumactive}
also brought a renaissance of mappings onto effective Fokker-Planck equations (FPEs) for the positional variable, dating back to a significant body of work done in the eighties \cite{BOOOK}.
Most notably, multicomponent versions~\cite{sharma2017,maggi2015sr,rein2016} of the approach by Fox~\cite{fox1,fox2}
and the Unified colored noise approximation (UCNA)~\cite{ucna1,ucna2}
have been adopted to study structural~\cite{activePair,fily2017,maggi2015sr,marconi2015,marconi2016sr,faragebrader2015,wittmannbrader2016,rein2016,martin2020,caprini2022spatial2},
 mechanical~\cite{activeCurvature,activePressure,marconi2016,marconiExactpressure2017}
 and dynamical \cite{sharma2017} properties of active fluids. 
Although the nature of these two approaches is quite different, 
they result in the same effective configurational probability distribution in the steady state.
However, a detailed comparison~\cite{activePair} revealed that Fox's result better reflects the behavior of active particles in the presence of an additional white noise, naturally emerging in a Brownian system.
Also for mixtures of two particles with different activity Fox's approach has been shown to be better reflect the nonequilibrium behavior \cite{activeMixture}.

The method by Fox belongs to a large class of approximation schemes providing an effective FPE for the configurational variables. 
A generic strategy to arrive at such a result involves writing down a formally exact FPE and subsequently eliminating the fluctuating variable,
 for which various methods have been exploited \cite{BOOOK,vanKampen1976,sanchogarrido1982cumulant,grigolini1986,luczka2005review}. 
It has been argued \cite{peacock1988,grigolini1986} that it is the particular choice of subsequent approximations
that determines the central quantity within the approximate FPE, i.e., the effective diffusivity.
The most common outcomes of these procedures 
either correspond to the so-called ``best Fokker-Planck approximation'' (BFPA) \cite{sanchogunton1982,peacock1988,grigolini1986,masoliver1987,lindenberg1983,lindenberg1983MC,lindenberg1984MC,parrondo1990GEOM} 
or are equivalent to the Fox approximation \cite{fox1,fox2,grigolini1986,parrondo1990GEOM,sharma2017,rein2016}.
Most notably, the Fox result can be recovered in two contrasting ways,
namely by performing a particular expansion of the BFPA to first order \cite{grigolini1986,masoliver1987}, as well as, by taking the derivation of the BFPA to higher orders \cite{faetti1987,grigolini1988,faetti1988}.
Among the approximation schemes toward an effective FPE, the UCNA takes a special role, since all approximations are made on the level of the governing stochastic Langevin differential equation~\cite{maggi2015sr,marconi2015}.
As a result, the dynamical behavior predicted by the UCNA is inherently different and only the steady-state results are directly comparable to other approaches~\cite{activePair}.

In contrast to Fox and UCNA, the BFPA has not been discussed in the contemporary context of active particles.
A possible reason for this is the decrease of its popularity following an extensive debate of whether even the one-component BFPA does justice to its auspicious name,
which can be recapitulated by considering the following three aspects.
First, a well-known problem of both the BFPA and the Fox approach is that the effective diffusivity is not always positive definite,
which may result in an ill-defined probability distribution.
However, the implicit form defined within the BFPA is even more uncontrollable than the analytic Fox expression,
 as the latter allows for a simple empirical modification to enforce positive definiteness \cite{activePair}.
  As an alternative rectification, a time-dependent approximation for the diffusion coefficient has been proposed \cite{tsironisDoft1,tsironisDoft2}. 
In general, it has been argued that any effective FPE 
possessing a positive definite diffusivity comes along with a significant increase in computational complexity \cite{venkatesh1993}. 
Second, the additional shortcomings specific to the BFPA have been revealed by addressing, e.g., 
empirical modifications \cite{grigolini1988},
 higher-order terms \cite{faetti1988,faetti1987,grigolini1988},
 details of the underlying model system \cite{peacock1989},
 alternative approximations in the derivation \cite{luczka1989} 
or regularization techniques \cite{hannibal1990}.
Finally, the major point of criticism of the BFPA arises from the fact that it is not exact in the limit of infinite persistence time of the colored noise,
in contrast to both Fox and UCNA, see Refs.~\onlinecite{faetti1988,grigolini1988} for a detailed discussion.
The higher appreciation of these latter approaches is thus based on their interpolation between two exact limits, as all effective FPEs are naturally exact for (passive) particles in the limit of uncorrelated (thermal) white noise.
For these reasons the BFPA is not the first method of choice in the recent applications of effective equilibrium descriptions to active particles.
While the name BFPA justifiably remained for historical reasons, the above insights have been summarized by Grigolini in the intentionally contradictory 
statement that ``an approximation to the best Fokker-Planck approximation, turns out to be better than the best Fokker-Planck approximation itself''~\cite{grigolini1988}.

Despite its ill-fated history, there are two reasons that make it worthwhile to resurrect the multicomponent BFPA in view of an application to AOUPs.
First, it allows to draw a connection to the recently obtained systematic small-activity expansions \cite{fodor2016,bonilla2019,martin2020},
which provide an accurate description of the stationary configurational probability distribution upon subsequently integrating out of the velocity coordinate and performing a resummation.
In contrast to Fox and UCNA, these calculations admit a dependence on the interaction potential beyond its second derivatives - just like the classical expansions providing the foundations of the BFPA \cite{grigolini1986,grigolini1988,sancho1980,sanchogunton1982,sanchogarrido1982cumulant}.
Relatedly, there exist some systems where the BFPA actually captures qualitative features that are not appreciated by the Fox theory, like, e.g., a nonlocal curvature dependence \cite{grigolini1986}.
Second, the explicit knowledge of the Langevin equations for the colored-noise system of interest is not necessarily required to derive
the BFPA, e.g., when using a projection-operator formalism \cite{moriPO,zwanzigPO,tevrugtPO}.
This opens up a new avenue to derive effective FPE models when the current methods to derive the Fox and UCNA for AOUPs cannot be readily employed.

One  such system, for which  the question on what is the proper form of the corresponding (overdamped) Langevin equation of motion is not easily answered \cite{vuijk2019magnetic},
is the Brownian motion of charged active particles evolving under the influence of the Lorentz force in a spatially inhomogeneous magnetic field \cite{vuijk2020magnetic}. 
The FPE for the overdamped motion under Lorentz force, however, can be derived without the explicit knowledge of the corresponding Langevin equation \cite{chun2018, vuijk2019magnetic}.
The diffusion coefficient that enters the FPE is tensorial and, notably, has an antisymmetric part. 
This apparent violation of the symmetry of the diffusion tensor has its roots in the broken time-reversal symmetry due to the Lorentz force. 
With such a diffusion tensor, density gradients give rise to fluxes both along and perpendicular to them \cite{abdoli2020fluxes}. 
These dynamics, which basically represent the diffusive version of the classical Hall effect, 
are characteristic of a class of systems referred to as \textit{odd-diffusive}, which have recently received significant attention \cite{abdoli2020thermostat, abdoli2020reset, abdoli2022resetactive, hargus2021odd, yasuda2022timecorrelations, fruchart2023odd, hargus2024theflux, kalz2024field}.
Odd-diffusive systems can exhibit unusual or counter-intuitive behavior,
such as superballistic motion of overdamped Rouse dimers \cite{shinde2022strongly}, 
 interaction-enhanced self-diffusion \cite{kalz2022collisions} 
or oscillatory autocorrelations \cite{kalz2024oscillatory}.
Moreover, the dynamics of interacting particles under Lorentz force show intriguing similarity to 
systems dominated by a Magnus force \cite{reichhardt2020dynamics,reichhardt2022active}
and point vortices in superfluids \cite{doshi2021vortices} or liquid systems~\cite{aref1999four,tophoj2008chaotic}.

In this paper, we use projection operators to 
derive a general FPE in the form of the multicomponent BFPA for charged AOUPs subject to a Lorentz force in an inhomogeneous magnetic field.
To this end, we discuss in Sec.~\ref{sec_theory}  the AOUP model and its approximate description in terms of effective FPEs, focusing on the BFPA and its expansions.
Then we study in Sec.~\ref{sec_Lorentz} the behavior of overdamped AOUPs in the presence of homogeneous and inhomogeneous magnetic fields,
using numerical simulations and the generalized Fox formula obtained in the first iteration step of our general BFPA result. 
We conclude in Sec.~\ref{sec_conclusion} and discuss possible future applications.

\section{Effective Fokker-Planck equations \label{sec_theory}}

 Our first goal is to introduce the BFPA and understand its relation to other available approaches.
For the moment, we thus consider interacting AOUPs in the absence of a magnetic field and recapitulate the current status of effective FPEs.

\subsection{The AOUP model}

 In our equations of motion, we use a component-wise notation~\cite{marconi2015} for the $\di N$ coordinates $x_\alpha$ of $N$ AOUPs in $\di$ dimensions and indicate the set of all coordinates $\{x_\alpha\}$ (or other variables) for $\alpha=1,\ldots,\di N$ by curly brackets.
Then, the Langevin equations of the AOUPs read
\begin{equation}
\dot{x}_\alpha(t) = \gamma^{-1}F_\alpha(\{x_\beta\}) + I_\text{t}\xi_\alpha(t) + \chi_\alpha(t)\,,
\label{eq_AOUPs}
\end{equation}
where the stationary stochastic processes $\chi_\alpha(t)$ evolve in time according to
\begin{equation}
\dot{\chi}_\alpha(t)=-\frac{\chi_\alpha(t)}{\taua}+\frac{\eta_\alpha(t)}{\taua}\,.
\label{eq_AOUPsDEF}
\end{equation}
The Gaussian white noises $\xi_\alpha$ and $\eta_\alpha$ have the correlators
$\langle\xi_\alpha(t)\xi_\beta(t')\rangle\!=\!2D_\text{t}\delta_{\alpha\beta}\delta(t-t')$ and
$\langle\eta_\alpha(t)\eta_\beta(t')\rangle\!=\!2D_\text{a}\delta_{\alpha\beta}\delta(t-t')$,
where $D_\text{t}$ and $D_\text{a}$ are the diffusion coefficients characterizing passive Brownian motion and the active propulsion, respectively.
Moreover, $\taua$ is the persistence time of the active motion, $\gamma=(\beta D_\text{t})^{-1}$ is the friction coefficient satisfying the fluctuation-dissipation theorem with the inverse temperature $\beta$, and $I_\text{t}\in\{0,1\}$ serves as a characteristic function to specify whether or not
the (translational) Brownian noise $\xi_\alpha(t)$ is present.
Finally, $F_\alpha$ are the conservative forces on $x_\alpha$, which introduce a typical unit length scale $d$.

The Ornstein-Uhlenbeck processes, Eq.~\eqref{eq_AOUPsDEF}, evolve independently of the spatial coordinates.
They are Gaussian processes with zero mean and the correlation
\begin{equation}
 \langle\chi_\alpha(t)\chi_\beta(t')\rangle
 =\frac{D_\text{a}}{\taua}\delta_{\alpha\beta}e^{-\frac{|t\!-\!t'|}{\taua}}\,.
 \label{eq_AOUPsCORR}
\end{equation}
The steady-state probability distribution
\begin{align}
R_N(\{\chi_\alpha\})=\left(\frac{\taua}{2\pi D_\text{a}}\right)^{\!\frac{\di N}{2}}\prod_\alpha\exp\left(-\frac{\taua}{2D_\text{a}}\chi_\alpha^2\right)
 %=\prod_\alpha\rho(\chi_\alpha)\,,
 \label{eq_RN}
\end{align}
 of the associated variables $\{\chi_\alpha\}$ is a product of individual Gaussians $\rho(\chi)=\sqrt{\taua/(2\pi D_\text{a})}\exp\left(-\taua \chi^2/(2D_\text{a})\right)$.
The rotational motion of active Brownian particles with self-propulsion velocity $v_0$ can be approximately represented by AOUPs \cite{faragebrader2015,activePair} 
when identifying $\taua$ with the timescale for rotational Brownian motion, setting $D_\text{a}=v_0^2\taua/\di$
and neglecting higher-order correlations of the active force.
In appendix~\ref{app_numerics},  we describe our numerical simulations of the AOUP model, 
including a generalization to include the Lorentz force on charged AOUPs in a magnetic field.

\subsection{The Fox approach recapitulated}

Here, we briefly state the results from the literature for an effecitve Fokker-Planck equation for the configurational probability distribution $f_N(\{x_\alpha\},t)$
of the particle positions, representing the stochastic process specified by Eq.~\eqref{eq_AOUPs}.
The multicomponent Fox approximation \cite{sharma2017,rein2016,parrondo1990GEOM} has the general form
\begin{equation}
\frac{\partial_tf_N}{D_\text{t}}=-\partial_\alpha\beta F_\alpha 
f_N+\partial_\alpha\partial_\beta(\mathcal{D}_{\alpha\beta}f_N)\,,
\label{eq_FPEgen}
\end{equation}
where $\mathcal{D}_{\alpha\beta}$ is a dimensionless effective diffusion tensor and $\partial_\alpha=\partial/\partial x_\alpha$
denotes the partial derivative in space with respect to the component $x_\alpha$ of the particle position 
(which is not to be confused with the stochastic process $x_\alpha(t)$ in Eq.~\eqref{eq_AOUPs}). 
We use the convention that repeated indices are summed over from 1 to $\di N$.
 The notion of $\mathcal{D}_{\alpha\beta}$ as a diffusion tensor becomes evident from rearranging Eq.~\eqref{eq_FPEgen} 
in the form of a diffusion equation with an effective force \cite{activePair}.

The components of the effective diffusion tensor are given by
\begin{align}
\mathcal{D}^\FOX_{\alpha\beta}(\{x_\gamma\})=I_\text{t}\delta_{\alpha\beta}+\Da\,\Gamma^{-1}_{\alpha\beta}\,,\label{eq_GammaF}
\end{align}
with the dimensionless active diffusivity 
$\Da=D_\text{a}/D_\text{t}$ and the mobility matrix 
\begin{align} \label{eq_Gamma}
 \Gamma_{\alpha\beta}(\{x_\gamma\})=\delta_{\alpha\beta}-\taua D_\text{t}\beta\partial_\beta F_\alpha\,.
\end{align}
 Here and throughout the manuscript $\Gamma^{-1}_{\alpha\beta}$ denotes the components of the inverse matrix $\boldsymbol{\Gamma}^{-1}$ and not the inverse $1/\Gamma_{\alpha\beta}$ of a single component.
For completeness, the effective diffusion tensor in the UCNA reads
\begin{align}
\mathcal{D}^\UCNA_{\alpha\beta}(\{x_\gamma\})&=(I_\text{t}+\Da)\,\Gamma^{-1}_{\alpha\beta}\,, \label{eq_GammaU}
\end{align}
such that we find $\mathcal{D}^\UCNA_{\alpha\beta}=\mathcal{D}^\FOX_{\alpha\beta}$ in the absence of thermal noise ($I_\text{t}=0$).
However, in the UCNA, the effective FPE generally differs from Eq.~\eqref{eq_FPEgen} by a factor $\Gamma_{\alpha\beta}$,
see Ref.~\cite{activePair} for a detailed comparison of these two approaches.

\subsection{Derivation of the BFPA \label{sec_deriveBFPA}}

As a next step, we extend the number of available effective configurational FPEs by deriving the
multicomponent BFPA~\cite{lindenberg1984MC} for AOUPs (see also appendix~\ref{app_derivation}). To this end, let us first
consider the FPE
\begin{equation}
\partial_tP_N=(\mathcal{L}_a+\mathcal{L}_b+\mathcal{L}_1)P_N\,,
\label{eq_FPEjointLi}
\end{equation}
for the joint probability distribution $P_N(\{x_\alpha\},\{\chi_\alpha\},t)$ 
associated with Eqs.~\eqref{eq_AOUPs} and~\eqref{eq_AOUPsDEF},
where
\begin{align}
\mathcal{L}_a&:=-D_\text{t}\partial_\alpha(\beta F_\alpha-I_\text{t}\partial_\alpha)\,, \label{eq_La}\\
\mathcal{L}_b&:=\frac{1}{\taua}\frac{\partial}{\partial\chi_\alpha}\chi_\alpha+\frac{D_\text{a}}{\taua^2}\frac{\partial}{\partial\chi_\alpha}\frac{\partial}{\partial\chi_\alpha}\,, \label{eq_Lb}\\
\mathcal{L}_1
&:=-\chi_\alpha \partial_\alpha \,.
\label{eq_L1}
\end{align}
The motivation for this splitting into three scalar operators is that
the variables $\{\chi_\alpha\}$ representing the (correlated) active fluctuations
 are not of direct physical interest.
Due to the cross-terms in the interaction operator $\mathcal{L}_1$, a simple factorization of
$P_N(\{x_\alpha\},\{\chi_\alpha\},t)$ into probability distributions of the type $f_N(\{x_\alpha\},t)$ and $R_N(\{\chi_\alpha\},t)$ is not possible.
  We thus seek to identify an approximate FPE for the marginal probability distribution $f_N(\{x_\alpha\},t)$ in the form of Eq.~\eqref{eq_FPEgen} by eliminating the dependence on $\{\chi_\alpha\}$.

We proceed by projecting Eq.~\eqref{eq_FPEjointLi} onto the (multivariate) steady-state distribution $R_N(\{\chi_\alpha\})$, given by Eq.~\eqref{eq_RN}, 
which solves $\mathcal{L}_b\,R_N=0$.
As detailed in appendix~\ref{sec_PO}, this procedure yields the formal expression 
\begin{equation}
\partial_tf_N=
\mathcal{L}_af_N+\int_0^t\upd s\left\langle \mathcal{L}_\chi(0)e^{\mathcal{L}_as}\mathcal{L}_\chi(s)e^{-\mathcal{L}_as}\right\rangle f_N(t)\!\!\!
\label{eq_FPEformal}
\end{equation}
for the configurational probability distribution at second perturbation order, valid up to linear order in $D_\text{a}\taua$,
where the angle brackets denote an average with respect to $R_N$.
Here, we defined the dynamical evolution operator $\mathcal{L}_\chi(t):=-\chi_\alpha(t)\partial_\alpha$,
which follows from $\mathcal{L}_1$ in Eq.~\eqref{eq_L1}
upon replacing the random variables $\{\chi_\alpha\}$ with the explicit time-dependent Ornstein-Uhlenbeck processes $\{\chi_\alpha(t)\}$.

As detailed in appendix \ref{sec_BFPAmulti}, the term under the integral in Eq.~\eqref{eq_FPEformal} 
can be rewritten in terms of the exponential correlator, Eq.~\eqref{eq_AOUPsCORR}, of the Ornstein-Uhlenbeck noise
and a tensorial quantity depending on spatial derivatives of $F_\alpha$. 
Carrying out the integral and approximating the result as a time-independent diffusion tensor in appendix \ref{sec_BFPAmulti2},
we recast Eq.~\eqref{eq_FPEformal} in the desired form of Eq.~\eqref{eq_FPEgen}.
The resulting generalized diffusion tensor reads
\begin{align}
\mathcal{D}^\BFPA_{\alpha\beta}(\{x_\gamma\})=I_\text{t}\delta_{\alpha\beta}+\mathcal{D}_{\alpha\beta}\,,\label{eq_GammaB}
\end{align}
 where the active part $\mathcal{D}_{\alpha\beta}=\Da \sum_{n=0}^\infty\tilde{\mathcal{D}}^{(n)}_{\gamma\beta}$
can be obtained from an infinite series with the coefficients
\begin{align}
 \tilde{\mathcal{D}}^{(n)}_{\alpha\beta}=\taua D_\text{t}\beta\left(\tilde{\mathcal{D}}^{(n-1)}_{\gamma\beta} \partial_\gamma F_\alpha - F_\gamma\partial_\gamma\tilde{\mathcal{D}}^{(n-1)}_{\alpha\beta}\right)\!\!\!
\label{eq_DeffALLGsum}
\end{align}
for $n\geq1$ and the zero-order term $\tilde{\mathcal{D}}^{(0)}_{\alpha\beta} =\delta_{\alpha\beta}$.

  Alternatively, we also show in appendix \ref{sec_BFPAmulti2} that $\mathcal{D}_{\alpha\beta}$
solves the differential equation
\begin{align}
\mathcal{D}_{\alpha\beta}\,\Gamma_{\gamma\alpha}&=\Da \delta_{\gamma\beta}-
\taua D_\text{t} \beta F_\alpha\partial_\alpha\mathcal{D}_{\gamma\beta}
\label{eq_D23b2LORENTZ}
\end{align}
with the mobility matrix $\Gamma_{\alpha\beta}$ given by Eq.~\eqref{eq_Gamma}. 
The zero-order term (neglecting the derivative $\partial_\alpha\mathcal{D}_{\gamma\beta}$ in the second term on the right-hand side)
of the iterative solution of Eq.~\eqref{eq_D23b2LORENTZ} reads $\mathcal{D}^{(0)}_{\alpha\beta}=\Da\,\Gamma^{-1}_{\alpha\beta}$,
which means that we recover the Fox result $\mathcal{D}^\FOX_{\alpha\beta}(\{x_\gamma\})\equiv I_\text{t}\delta_{\alpha\beta}+\mathcal{D}^{(0)}_{\alpha\beta}$
from Eq.~\eqref{eq_GammaF} as a fundamental limiting case \cite{grigolini1986}.

\subsection{Expansions and results in one dimension \label{sec_1dBFPA}}

To illustrate the basic predictions of the BFPA, we set $I_\text{t}=0$ for the moment and recapitulate the most simplistic scenario of a single particle in a (dimensionless) one-dimensional external potential $\phi(x)$, such that $\beta F(x)=-\phi'(x)$.
Here and in the following, we use the prime to denote the derivative of any function $a(x)$ with respect to the single coordinate $x\equiv x_1$, i.e., $a'(x)\equiv \partial_x a(x)$. 
Then, Eq.~\eqref{eq_D23b2LORENTZ} reduces to
\begin{align}
\mathcal{D}(x)(1+\taua D_\text{t}\phi''(x))&=\Da +\taua D_\text{t}\phi'(x)\mathcal{D}'(x)\,,
\label{eq_D23_1d}
\end{align}
where $\mathcal{D}\equiv\mathcal{D}_{11}$. 
When solving this differential equation, the integration constant should be chosen such that $\mathcal{D}(x)$
  takes the bulk value $\Da$ whenever $\phi(x)=0$ and $\mathcal{D}(x)=0$ in the limit that the potential $\phi(x)$ becomes infinitely high.
Recall that setting $\mathcal{D}'(x)=0$ returns the Fox approximation $\mathcal{D}^{(0)}(x)$ 
which is known to be consistent with these conditions.
Starting from this leading expression the iterative solution of Eq.~\eqref{eq_D23_1d} becomes
\begin{align}
\mathcal{D}^{(n+1)}(x)&=\frac{\Da +\taua D_\text{t}\phi'(x)\left(\mathcal{D}^{(n)}\right)'(x)}{(1+\taua D_\text{t}\phi''(x))}\,
\label{eq_D23_1dit}
\end{align}
for $n\geq0$.
A convenient way to compare different theories for effective FPEs is to calculate the steady-state distribution $f_N(x)\propto\exp(-\phi_\text{eff}(x))$ 
or, equivalently, the effective potential
 $\phi_\text{eff}(x)
=\int_{x_0}^x\frac{\phi'(y)}{\mathcal{D}(y)}\upd y+\ln(\mathcal{D}(x))$, where $x_0$ is a conveniently chosen reference position such that $\phi(x_0)=0$.
This result is obtained by setting $\partial_tf_N=0$ in the one-dimensional version
of Eq.~\eqref{eq_FPEgen}, which yields the \textit{effective equilibrium} condition
\begin{equation}
0=\partial_x\phi'(x)
f_N(x)+\partial_x^2\,\mathcal{D}(x)f_N(x)\,.
\label{eq_FPEgen1d}
\end{equation}

To better understand the difference between the different results for the effective potentials
it is insightful to look at the expansions of the analytic formulas at short persistence time $\taua$.
The small-$\taua$ expansion of the Fox theory yields
\begin{align}
\phi^\FOX_\text{eff}(x)=\frac{\phi}{\Da}-\taua D_\text{t}\left(\phi''-\frac{(\phi')^2}{2\Da}\right)+\frac{\taua^2D_\text{t}^2(\phi'')^2}{2}+\mathcal{O}(\taua^3)
\end{align}
up to quadratic order.
This expansion reflects a characteristic feature of the full Fox expression, namely that it contains only first and second derivatives of the bare potential.
In contrast, the effective potential
\begin{align}
\phi^\BFPA_\text{eff}(x)=\phi^\FOX_\text{eff}-\taua^2D_\text{t}^2\left(\int_{x_0}^x\frac{(\phi')^2\phi'''}{2\Da}\upd y-\phi'\phi'''\right)+\mathcal{O}(\taua^3)
\label{eq_phieffBFPA}
\end{align}
of the BFPA 
differs at quadratic order in $\taua$ from the Fox result, as it contains also third derivatives of the bare potential $\phi(x)$.

One might argue that the inclusion of all higher-order derivatives should constitute an advantage of the BFPA.
For example, while Fox theory can be used \cite{activeCurvature} to study the pressure of active particles at curved walls \cite{smallenburg2015}, it fails to accurately reflect nonlocal effects in curvature-dependent properties \cite{nikola2016,sandford2018}.
 A related one-dimensional paradigmatic example of a periodic 
 potential $\phi(x)=-\sin(\omega x)$ with constant wave number $\omega$ was first discussed by Grigolini in Ref.~\onlinecite{grigolini1986}.
 The analytic results 
\begin{align}
\mathcal{D}^{(0)}(x)&= \frac{\Da}{1+\omega^2\taua D_\text{t}\sin(\omega x)}\,, \label{eq_sinmodFOX}\\
\mathcal{D}(x)&= \Da\frac{1-\omega^2\taua D_\text{t}\sin(\omega x)}{1-\omega^4\taua^2 D_\text{t}^2}\,\label{eq_sinmodBFPA}
\end{align}
underline that the diffusivity $\mathcal{D}^{(0)}(x)$ in the Fox approach is equal to the bulk value $\Da$ whenever the bare potential has zero curvature, 
whereas the BFPA solution $\mathcal{D}(x)$ admits a nonlocal curvature dependence.

\begin{figure}
\includegraphics[width=0.5\textwidth] {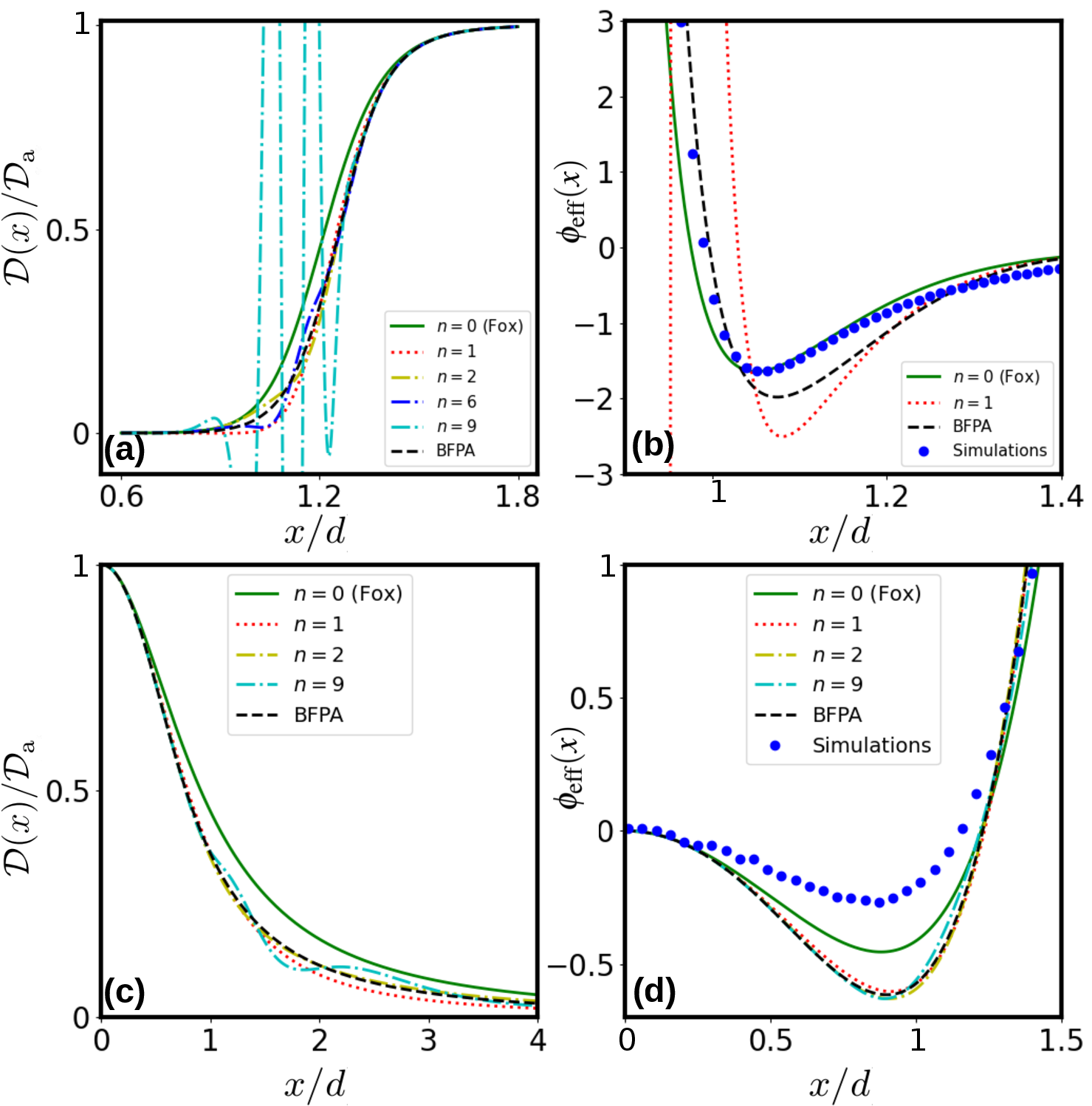}

\vspace*{-.1cm}

\caption{Comparison of the BFPA (dashed line) to its iterative solution (lines including dots, as labeled) including the Fox result (solid line) at second order.
We consider a single particle without thermal white noise ($I_\text{t}=0$) in an external potential $\phi(x)$ in one spatial dimension, where $\tau=0.1d^2/D_\text{t}$ and $\Da=4.8$.
\textbf{(a)} Effective diffusivity $\mathcal{D}(x)$ for a soft-repulsive potential $\phi(x)=(x/d)^{-12}$.
\textbf{(b)} Effective potential $\phi_\text{eff}(x)$ for a soft-repulsive potential $\phi(x)=(x/d)^{-12}$. Symbols denote the results of AOUP simulations.
\textbf{(c)} Effective diffusivity $\mathcal{D}(x)$  for a soft trap $\phi(x)=(x/d)^{4}$.
\textbf{(d)} Effective potential $\phi_\text{eff}(x)$ for a soft trap $\phi(x)=(x/d)^{4}$. Symbols denote the results of AOUP simulations.
\label{fig_1d}
}
\end{figure}

In a harmonic potential $\phi(x)=(x/d)^2$ all derivatives of $\phi$ of order higher than $\phi''$ vanish, 
such that the BFPA becomes identical to the Fox approach.
 Two other representative one-dimensional potentials are considered in Fig.~\ref{fig_1d} to test the iterative solution of the BFPA and learn more about the difference to the Fox approximation at zero order. 
For the soft-repulsive potential $\phi(x)=(x/d)^{-12}$, 
 commonly used to model the soft repulsion of a wall or between two interacting particles~\cite{faragebrader2015,wittmannbrader2016,activePair,maggi2015sr},
 the diffusivity of both approaches exhibits a qualitatively similar behavior, but the transition from the value 0 to the bulk value $\Da$ occurs for higher values of $x$ in the BFPA, compare Fig.~\ref{fig_1d}a.
 A related shift of the attractive region of the corresponding effective potential can be seen in Fig.~\ref{fig_1d}b.
 Comparing to numerical simulations, we see that this prediction of the BFPA is in fact quantitatively inaccurate.
  Regarding the iterative solution, already including the first correction term $\mathcal{D}^{(1)}$ 
  to the Fox term $\mathcal{D}^{(0)}$ results in a fairly good agreement 
  with the full solution $\mathcal{D}$ of the BFPA.
However, such an expansion turns out to be unstable,
as can be clearly seen from adding further terms of higher order.
 Most notably, there is an unphysical divergence of 
 the effective potential $\phi_\text{eff}(x)$ already at first order.
Similar qualitative conclusions can be drawn from a potential $\phi(x)=(x/d)^{4}$, providing a simple model for a soft particle trap \cite{activeCurvature,marconi2015}.
Here, the first terms of the iterative solution of the BFPA provides a decent approximation for the full result, compare Figs.~\ref{fig_1d}c and d
but it becomes apparent from the oscillations emerging in the higher-order terms that a full resummation would not converge even for this relatively simple potential with a finite number of nonvanishing derivatives.
 As the agreement with numerical simulations is again better for the Fox approach, 
we conclude that, even in one spatial dimension, the usefulness of the BFPA in the context of active particles is limited,
in particular, when taking into account to the increased complexity of its evaluation.
 In higher dimensions, we show in appendix \ref{sec_results2Cnew} 
 that the problems of the BFPA become even more dramatic:
while the issues outlined for a one-dimensional system persist for the radial Eigenvalue of the effective diffusion tensor,
the differential equation for the polar Eigenvalue has only a finite region of support
and thus cannot be determined unambiguously.

\subsection{Comparison with exact expansions}

 Finally, we establish a connection to other perturbative expansions recently developed for AOUPs~\cite{fodor2016,martin2020,bonilla2019,marconi2017_expansion,martin2021_expansion,caprini2022spatial2}, where we stick here to the simplified picture of a single particle in one spatial dimension, as in Sec.~\ref{sec_1dBFPA}.
 The common starting point of these approaches is a change of variables
 to replace the stochastic process $\chi$ 
 with the velocity $v(t):=\gamma^{-1}F(x) +\chi(t)$,
implicitly specified by Eq.~\eqref{eq_AOUPs} with $I_\text{t}=0$,
 which transforms the equation of motion into an effectively underdamped equation only containing the white noise $\eta(t)$ as in Eq.~\eqref{eq_AOUPsDEF}.
The corresponding FPE can be solved perturbatively, in terms of the small parameter $\taua$ (as in the derivation of the BFPA), 
and the configurational probability distribution can be eventually obtained by integrating out the velocities.

In such a formally exact small-$\taua$ expansion, there also emerge so-called nondiffusive terms in the effective diffusivity $\mathcal{D}(x)$ beyond the linear order in $\taua$, i.e., there are expressions proportional to $\Da^n$ with $n>1$.
Driven by the goal to obtain a closed theory accounting for all orders in $\taua$, 
these terms have have often been removed by introducing approximations~\cite{grigolini1986}.
 Examples for such approximations are those leading to the BFPA, i.e., not going beyond the second perturbation order in Eq.~\eqref{eq_FPEformal}, and the \textit{local linearization} at higher perturbation order, which is justified by recovering the Fox result as an extension to the BFPA \cite{faetti1988,faetti1987,grigolini1988}.

Specifically, taking the full perturbative expansion up to quadratic order in $\taua$, the exact effective potential~\cite{martin2020}
\begin{align}
\phi_\text{eff}^{(\text{p})}(x)=\phi^\BFPA_\text{eff}-\taua^2D_\text{t}^2\frac{\Da\phi''''}{2}+\mathcal{O}(\taua^3)
\label{eq_phieffPERT}
\end{align}
differs only by a single nondiffusive term from the BFPA result, Eq.~\eqref{eq_phieffBFPA}.
Including higher-order terms in this systematic expansion,
similar oscillations are found as in the expansions, Eq.~\eqref{eq_D23_1dit} or Eq.~\eqref{eq_phieffBFPA}, of the BFPA (cf.\ Fig.~\ref{fig_1d}a),
whereas a Borel resummation of the exact expansion, Eq.~\eqref{eq_phieffPERT}, overcomes this problem and improves upon both the BFPA and the Fox (UCNA) result for the effective potential considered in Fig.~\ref{fig_1d}d~\cite{martin2020,martin2021_expansion}.
A further comparison between the different approaches would be illuminating,
in particular in higher spatial dimensions.

\section{Overdamped AOUPs with Lorentz force \label{sec_Lorentz}}

Our next goal is an effective description of overdamped AOUPs carrying the charge $q$ in a space-dependent magnetic field $\bvec{B}(\bvec{r})$.
Since the action of the Lorentz force depends on the particle velocity, compare appendix~\ref{app_numerics}, it is not directly obvious how to describe the overdamped dynamics.
In the Fokker-Planck picture, this can be conveniently achieved by condensing the Lorentz force into a nonsymmetric diffusion tensor \cite{vuijk2019magnetic}, which means that the dynamics under Lorentz force are odd diffusive.
Moreover, given this starting point, it is possible to directly derive an effective FPE along the lines of Sec.~\ref{sec_deriveBFPA}, which we outline and analyze below.

\subsection{General  FPE}

For simplicity, we assume that  $\bvec{B}(\bvec{r})$ is oriented in $z$-direction and define the dimensionless diffusive Hall parameter $\kappa(\bvec{r})=\gamma^{-1}qB(\bvec{r})$.
Then the contribution of the magnetic field is contained within a tensor $G$ \cite{vuijk2020magnetic}, which we generalize to multiple particles as follows: 
\begin{align}
G_{\alpha\beta}(\{x_\gamma\})=\delta_{mn}\left(\delta_{ij}-\frac{\kappa}{1+\kappa^2}M_{ij}+\frac{\kappa^2}{1+\kappa^2}M_{ik}M_{kj}\right).
\label{eq_Gmag}
\end{align} 
Here, the indices $m,n\in\{1,\ldots,N\}$ are particle labels 
and $i,j,k\in\{1,\ldots,\mathfrak{d}\}$ denote the components in Cartesian coordinates, such that $\alpha=i+\mathfrak{d}(m-1)$ and $\beta=j+\mathfrak{d}(n-1)$.
For each particle, the matrices $M_{ij}=\epsilon_{ji3}$ can be written in terms of the Levi-Civita symbol $\epsilon_{ijk}$.

In analogy to the case of active Brownian particles \cite{vuijk2020magnetic}, the joint FPE 
\begin{align}
\partial_tP_N=&-\partial_\alpha G_{\alpha\beta}(D_\text{t}\beta F_\beta +\chi_\beta -I_\text{t}D_\text{t}\partial_\beta)P_N\cr
&+\frac{1}{\taua}\frac{\partial}{\partial\chi_\alpha}\chi_\alpha P_N
+\frac{D_\text{a}}{\taua^2}\frac{\partial}{\partial\chi_\alpha}\frac{\partial}{\partial\chi_\alpha}P_N\,,
\label{eq_FPEjoint}
\end{align}
for $P_N(\{x_\alpha\},\{\chi_\alpha\},t)$, generalizing Eq.~\eqref{eq_FPEjointLi}, acquires
 $G_{\alpha\beta}$ as a multiplicative factor to the active velocity $\boldsymbol\chi$ in the overdamped limit.
This is our
starting point to derive an effective FPE for charged AOUPs in a magnetic field.

\subsection{Effective configurational FPE}

Repeating the steps outlined in Sec.~\ref{sec_deriveBFPA} in a slightly more general way, 
it is straightforward to derive from Eq.~\eqref{eq_FPEjoint} a FPE for the configurational probability distribution $f_N(\{x_\alpha\},t)$
 in the form
\begin{equation}
\frac{\partial_tf_N}{D_\text{t}}=-\partial_\alpha G_{\alpha\beta}\left(\beta F_\beta 
f_N- 
\partial_\gamma(\mathcal{D}_{\gamma\beta}f_N)\right) \,,
\label{eq_FPEgenLORENTZ}
\end{equation}
see appendix \ref{app_derivation} for the full derivation.
 The diffusion tensor now depends in general on the spatial derivatives of both $F_\alpha$ and $G_{\alpha\beta}$.
 Although we also obtained the more general BFPA for charged AOUPs in a magnetic field, 
 we are mainly interested in the zero order term, from which we obtain
 \begin{align}
\mathcal{D}^\FOX_{\alpha\beta}(\{x_\gamma\})=I_\text{t}\delta_{\alpha\beta}+\Da\,\Gamma^{-1}_{\alpha\gamma}\,G_{\gamma\beta}\,
\label{eq_GammaFmag}
\end{align}
with the generalized mobility matrix
\begin{align} \label{eq_GammaLORENTZ}
 \Gamma_{\alpha\beta}(\{x_\gamma\})=\delta_{\alpha\beta}-\taua D_\text{t}\beta\partial_\beta G_{\alpha\gamma}F_\gamma\,.
\end{align}
 From this result, we recover the result of the Fox approximation, Eq.~\eqref{eq_GammaF},
 as a fundamental limiting case for $\kappa\rightarrow0$, since this implies $G_{\alpha\beta}=\delta_{\alpha\beta}$.

 A first inspection of these approximate dynamical equations 
reveals that the effective mobility $\Gamma_{\alpha\beta}=\delta_{\alpha\beta}$ in Eq.~\eqref{eq_GammaLORENTZ} becomes trivial in the special case of a free active particle ($F_\alpha=0$) in a magnetic field.
Nevertheless, the nonequilibrium coupling of the activity and the Lorentz force is also directly evident in this case through the factor $G_{\gamma\beta}$ in Eq.~\eqref{eq_GammaFmag}.
As a consequence, the stationary distribution of active particles is modulated by the inhomogeneous magnetic field, which is consistent with the predictions in Ref.~\onlinecite{vuijk2020magnetic}.  
The effective FPE~\eqref{eq_FPEgenLORENTZ} can describe a much more general scenario with additional conservative forcing  ($F_\alpha\neq0$), since the off-diagonal terms of $G_{\alpha\beta}$, which are responsible for the Lorentz fluxes \cite{chun2018,abdoli2020fluxes,abdoli2020reset}, also contribute to $\Gamma_{\alpha\beta}$.
Our  theory thus allows us to study a broad range of inhomogeneous scenarios,
which we illustrate in the following by calculating appropriate effective interaction potentials and probability distributions.

\begin{figure*}[t]
\includegraphics[width=\textwidth] {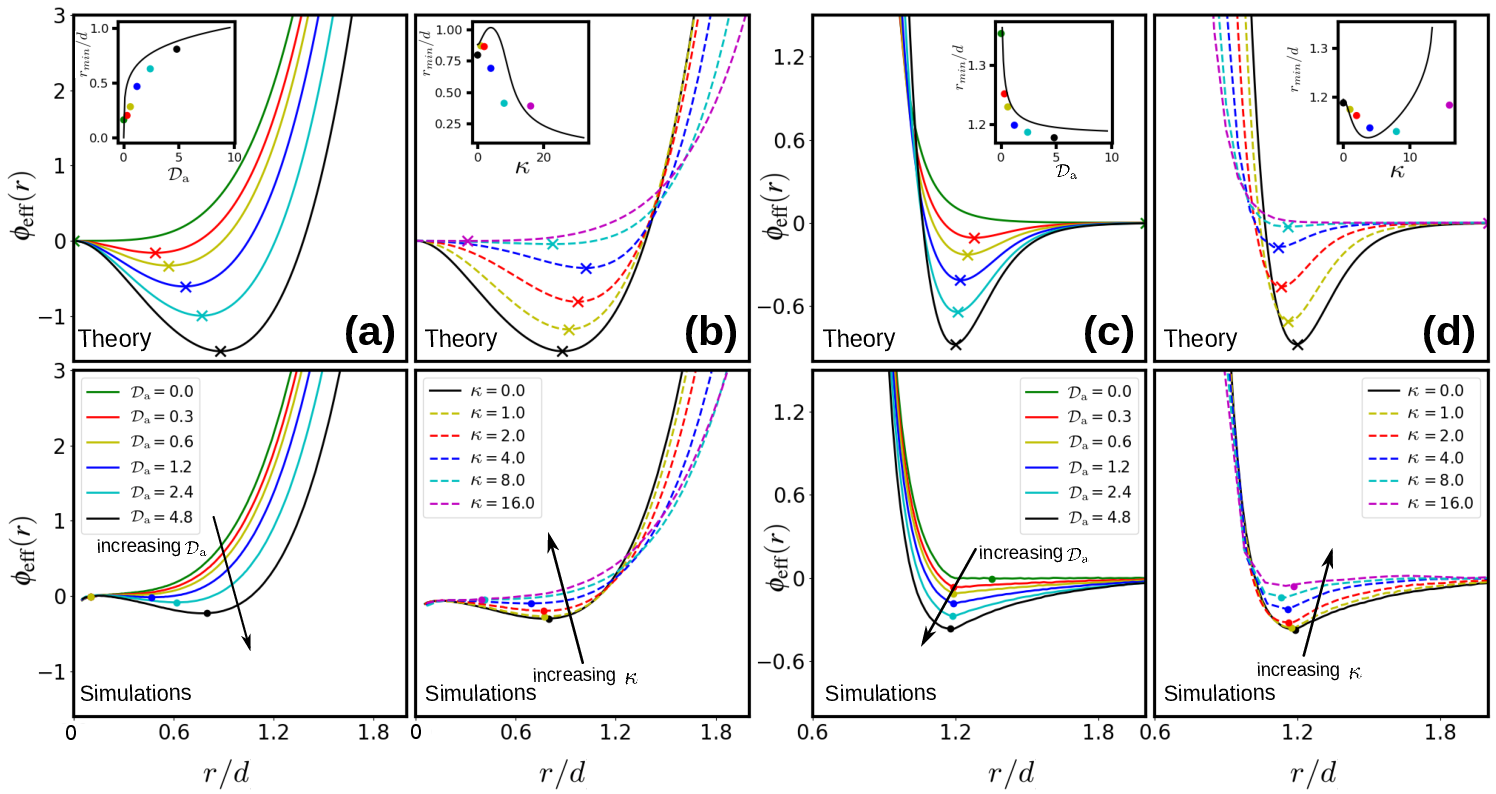}

\vspace*{-.1cm}

\caption{Effective potentials $\phi_\text{eff}(r)$ from theory (top row) compared to AOUP simulations (bottom row) in a radial geometry for a constant magnetic field with diffusive Hall parameter $\kappa$. The persistence time of the active motion is fixed to $\taua=0.5d^2/D_\text{t}$ and thermal white noise is present ($I_\text{t}=1$). 
We show the results for
\textbf{(a)} a soft trap with $\phi(r)=(r/d)^{4}$ for $\kappa=0$ and increasing active diffusivity $\Da$,
\textbf{(b)} a soft trap with $\phi(r)=(r/d)^{4}$ for fixed $\Da=4.8$ with increasing $\kappa$ and
\textbf{(c,d)} a soft-repulsive wall with $\phi(r)=(r/d)^{-12}$ for the same parameters.
The location of the potential minima is marked by crosses (theory) and dots (simulations), as well as, shown in the insets as a function of the parameters of interest using continuous lines (theory) and dots (simulations).
\label{fig_Bconst}
}
\end{figure*}

\subsection{Effective Lorentz potentials in two dimensions \label{sec_ELP2d}} 

To illustrate the predictions of Eq.~\eqref{eq_FPEgenLORENTZ},
we consider a single particle ($N=1$) in a two-dimensional external potential
acting perpendicular to the magnetic field.
In a polar geometry, 
the corresponding effective interaction potential $\phi_\text{eff}(r):=-\ln(f_1(r))+c$ (with $c$ being an arbitrary constant that is omitted in the following) can be obtained from
the stationary condition $\bvec{j}\cdot\hat{e}_r=0$ for the current
\begin{equation}
 \bvec{j} =- \boldsymbol{G} \cdot \left(\phi'f_1 \hat{e}_r+(f_1\boldsymbol{\mathcal{D}}^T)'\cdot\hat{e}_r+\frac{f_1}{r}(\partial_\varphi \boldsymbol{\mathcal{D}}^T)\cdot\hat{e}_\varphi\right).
 \label{eq_currentpolar}
\end{equation}
For convenience, we switched here from the Cartesian components with indices $\alpha,\beta\in\{1,2\}$ 
to a notation in terms of vectors and 2x2 matrices (lower and upper case bold letters, respectively)
represented in polar coordinates with the unit vectors 
$\hat{e}_r=(\cos\varphi,\sin\varphi)^T$ and $\hat{e}_\varphi=(-\sin\varphi,\cos\varphi)^T$, where the superscript $T$ indicates the transpose and the dash (which is short for $\partial_r$) and $\partial_\varphi$ denote the partial derivative with respect to $r$ and $\varphi$, respectively.
For further evaluation, we use the properties $(1+\kappa^2)\,\boldsymbol{G}\cdot \hat{e}_r=\hat{e}_r-\kappa\hat{e}_\varphi$
and $(1+\kappa^2)\,\boldsymbol{G}\cdot \hat{e}_\varphi=\kappa\hat{e}_r+\hat{e}_\varphi$ of the magnetic tensor $\boldsymbol{G}$ defined in Eq.~\eqref{eq_Gmag}.
As the effective diffusion tensor $\boldsymbol{\mathcal{D}}$ explicitly depends on $\boldsymbol{G}$, 
the unit vectors $\hat{e}_r$ and $\hat{e}_\varphi$ are no longer its Eigenvectors, in contrast to the special case $\kappa=0$ \cite{activePair}.
We therefore express 
$\boldsymbol{\mathcal{D}}=\sum_{i,j}\mathcal{D}_{ij}\hat{e}_i\otimes\hat{e}_j$
in a basis set consisting of these two unit vectors, where $i,j\in\{r,\varphi\}$ and $\otimes$ denotes the dyadic product.
Further defining $\mathcal{D}_{B}:=\mathcal{D}_{rr}+\kappa\mathcal{D}_{r\varphi}$, we find from the stationarity of Eq.~\eqref{eq_currentpolar} the general representation
\begin{equation}
\phi_\text{eff}'=\frac{\phi'}{\mathcal{D}_{B}}+\frac{\mathcal{D}_{B}'}{\mathcal{D}_{B}}+\frac{\mathcal{D}_{rr}-\mathcal{D}_{\varphi\varphi}-\kappa(\mathcal{D}_{r\varphi}+\mathcal{D}_{\varphi r})}{r\,\mathcal{D}_{B}}
\label{eq_phieffB}
\end{equation}
for the derivative of the effective potential $\phi_\text{eff}(r)$ in a polar geometry.

For a spatially constant magnetic field the relevant components of the diffusion tensor from Eq.~\eqref{eq_GammaFmag} read
\begin{align}
    \mathcal{D}_{B}&=I_\text{t}+\Da\,\frac{\kappa^2+E_1}{\kappa^2+E_1E_2}\,,\cr
    \mathcal{D}_{rr}&=I_\text{t}+\Da\,\frac{E_1}{\kappa^2+E_1E_2}\,,\cr
    \mathcal{D}_{\varphi\varphi}&=I_\text{t}+\Da\,\frac{E_2}{\kappa^2+E_1E_2}\,,
    \label{eq_Dcomponents}
\end{align}
while $\mathcal{D}_{\varphi r}=-\mathcal{D}_{r\varphi}$, such that these terms cancel in  Eq.~\eqref{eq_phieffB}.
 The full expressions for an arbitrary $\kappa(r)$ are given in appendix \ref{app_Dcomponents}.
The formulas in Eq.~\eqref{eq_Dcomponents} are conveniently expressed in terms of the Eigenvalues
\begin{align}
 E_2(r)&=1+\taua D_\text{t} \phi''(r)\,,\cr
 E_1(r)&=1+\taua D_\text{t} \frac{\phi'(r)}{r}
 \label{eq_EVs}
\end{align}
of the mobility matrix $\boldsymbol{\Gamma}$ in the absence of the magnetic field,
i.e., Eq.~\eqref{eq_Gamma} (or Eq.~\eqref{eq_GammaLORENTZ} with $\kappa=0$),
which correspond to the Eigenvectors $\hat{e}_r$ and $\hat{e}_\varphi$, respectively.
This notation allows us to directly connect to previous results for AOUPs in conservative external fields \cite{marconi2015,activeMixture,activePair,activeCurvature}
and, in the same spirit, remove possible divergences by empirically
setting $E_n\rightarrow1/(2-E_n)$ if $E_n<1$ for $n\in\{1,2\}$ \cite{activePair},
which we employ in the following for the polar Eigenvalue $E_1$ in the potential $\phi(r)=(r/d)^{-12}$.

In a planar geometry with a modulation of the external potential $\phi(x)$ in $x$-direction,
the effective potential $\phi_\text{eff}(x)$ can be deduced from Eq.~\eqref{eq_phieffB} by identifying $r$ and $\varphi$ with $x$ and $y$, respectively, while formally setting all terms with the explicit factor $1/r$ to zero  
implying $E_1\rightarrow1$ for the polar Eigenvalue, 
see appendix \ref{app_Dcomponents} for the explicit expressions.
The effective pair interaction between two charged particles in a magnetic field can be obtained in the same way and results from $\phi_\text{eff}$ upon setting $\taua\rightarrow2\taua$, just like for $\kappa=0$ \cite{activePair}.

\begin{figure*}[t]
\includegraphics[width=\textwidth] {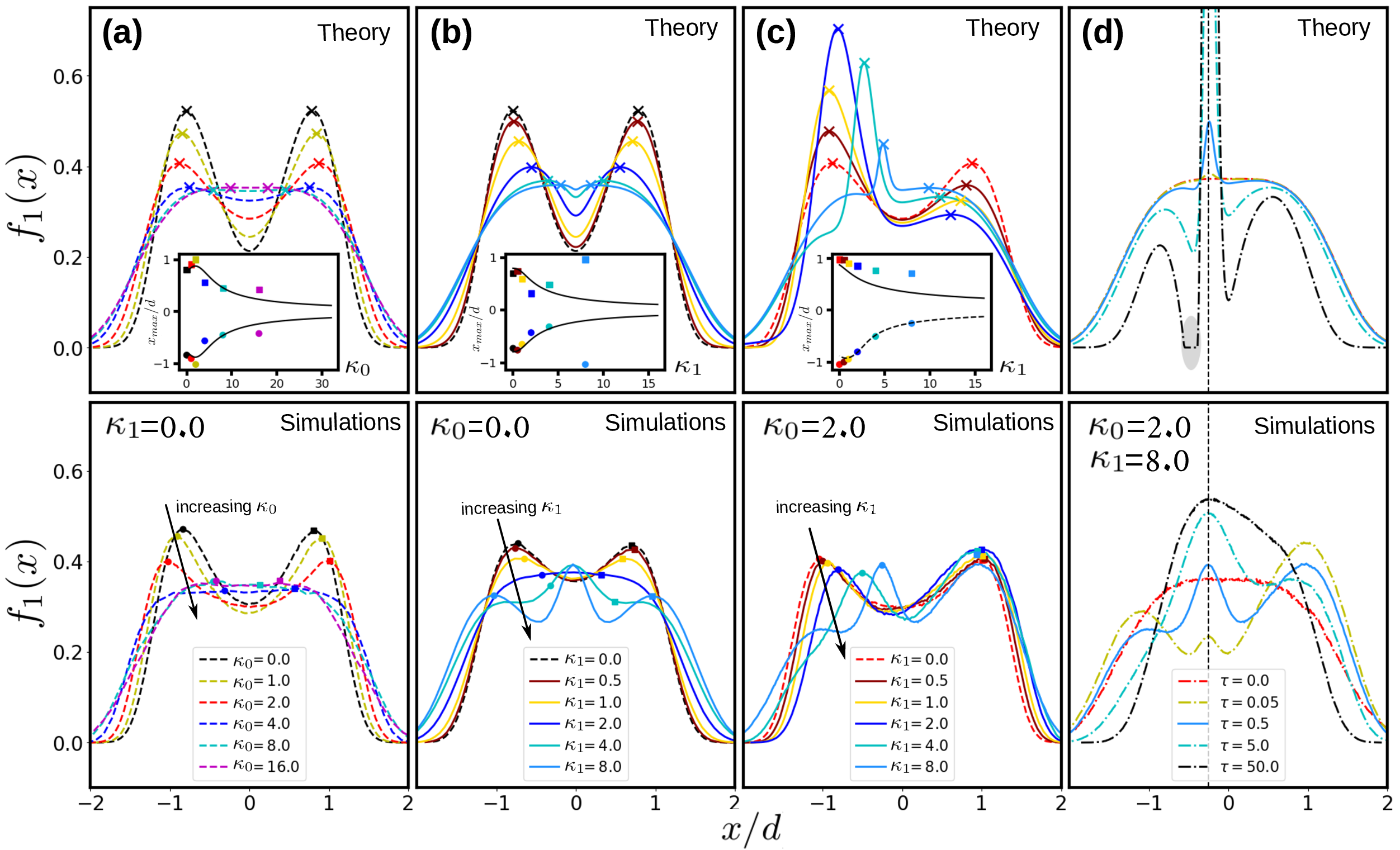}

\vspace*{-.1cm}

\caption{Normalized probability distributions $f_1(x)$ 
from theory (top row) compared to AOUPs simulations (bottom row) in a planar geometry with thermal white noise ($I_\text{t}=1$), related to the effective potentials as $\phi_\text{eff}=-\ln(f_1)$.
We show results for a soft trap with $\phi(x)=(x/d)^{4}$ and a spatially inhomogeneous magnetic field with a 
linear dependence of $\kappa(x)=\kappa_0+\kappa_1(x/d)$ on $x$. 
\textbf{(a,b,c)} Effect of changing the magnetic parameters $\kappa_0$ and $\kappa_1$ (see legend and annotations)
for fixed activity parameters $\taua=0.5d^2/D_\text{t}$ and $\Da=4.8$.
Analogously to Fig.~\ref{fig_Bconst}, the location of selected density maxima is marked by crosses (theory) and dots/squares (simulations), as well as, shown in the insets using continuous lines (theory) and according symbols (simulations).
\textbf{(d)} Effect of changing the persistence time $\taua$ for fixed $\kappa_0=2$, $\kappa_1=8$ and $\Da=4.8$ (see legend).
For the largest $\taua=50d^2/D_\text{t}$ the theory predicts a negative effective diffusivity in the shaded region for $-0.56>x/d>-0.38$, where $f_1$ is manually set to zero.
The vertical line marks the point where the magnetic field changes its sign, which closely coincides with the intermediate maxima of $f_1(x)$.
\label{fig_Bvar}
}
\end{figure*}

\subsection{Results for a constant magnetic field}

While the equilibrium distribution of (passive) particles is not affected by the action of the Lorentz force,
the steady-state properties of active particles can, in general, be altered, e.g., by a gradient of the magnetic field~\cite{vuijk2020magnetic}.
Here, we demonstrate that the effect of the Lorentz force on an active particle in an external potential can also be captured in an effective equilibrium picture.
Our theoretical results for the two representative repulsive external potentials $\phi(r)=(r/d)^{4}$ and $\phi(r)=(r/d)^{-12}$ are confirmed qualitatively by computer simulations, as shown in Fig.~\ref{fig_Bconst}.

As already known in the absence of a magnetic field, $\kappa=0$, the Fox approach predicts (Fig.~\ref{fig_Bconst}a,c) an increased effective attraction of the wall potential for an increasing activity~\cite{activePair}, reflecting the higher probability that active particles accumulate at the wall.
More specifically, we keep here the persistence time $\taua$ fixed and increase the active diffusivity $\Da$ and, along with it, the self-propulsion velocity $v_0$.
This results in a shift of the potential minimum in the direction of the wall, as rationalized by a force balance between the activity force (which is proportional to $v_0$) and the potential force (which increases when getting closer to the wall) \cite{caprini2022spatial2}.
The location of this minimum, which can be determined explicitly by setting the right-hand side of Eq.~\eqref{eq_phieffB} to zero, agrees particularly well with simulation results.

Keeping the activity fixed and considering $\kappa\neq0$, we predict (Fig.~\ref{fig_Bconst}b,d) that,
upon the action of the Lorentz force, the degree of effective attraction diminishes for both potentials.
Most intriguingly, the effective interaction becomes again purely repulsive when further increasing the strength of the magnetic field.
Regarding the location of the minima in the insets, we find that the Lorentz force due to a magnetic field of moderate strength ($\kappa\lesssim5$) allows the active particle to climb up further the potential gradient before the wall accumulation ceases at larger $\kappa$.
Moreover, our theory explicitly predicts that the effective interaction, as specified in Eq.~\eqref{eq_phieffB}, approaches a rescaled version 
\begin{equation}
    \lim_{\kappa\rightarrow\infty}
\phi_\text{eff}(r)=\frac{\phi(r)}{I_\text{t}+\Da}
\label{eq_limit}
\end{equation}
of the bare potential when taking the limit $\kappa\rightarrow\infty$ of an infinite magnetic field,
which is the same as in the limit $\tau\rightarrow0$ of vanishing persistence time.
In particular, this implies that the effective attraction fully disappears for sufficiently large values of $\kappa$.
This can also be intuitively understood by recognizing that the Lorentz force curves the particle's trajectory and thus continually turns its velocity direction, which impedes the persistent motion.
Hence, the persistence time $\taua$ of the stochastic active force is effectively reduced,
while its strength, represented by the active diffusivity $\Da$ in Eq.~\eqref{eq_limit}, remains the same.
We also observe such a trend when increasing $\kappa$ in our simulations, as the effective potentials become completely repulsive (Fig.~\ref{fig_Bconst}b,d),
but the limit $\kappa\rightarrow\infty$ cannot be explored numerically because of the finite time step.

\subsection{Results for an inhomogeneous magnetic field}
 
 As a next step, we are interested in a situation where the magnetic field is not spatially constant,
where we focus on the example of a two-dimensional system with a planar geometry for simplicity.
 Specifically, we consider the symmetric trapping potential $\phi(x)=(x/d)^{4}$ with the linearly increasing diffusive Hall parameter $\kappa(x)=\kappa_0+\kappa_1(x/d)$ along the Cartesian coordinate $x$, such that there is no spatial modulation along the $y$-direction.
 Crucially, $\kappa(x)$ changes its sign at $x/d=-\kappa_0/\kappa_1$.
 While the theory manages to provide a basic qualitative picture, it is not always accurate for all parameters, as becomes apparent from comparing the probability distributions for the particle location in Fig.~\ref{fig_Bvar}.
Hence, we first discuss the simulations to understand the basic physics of confined AOUPs with spatially dependent Lorentz force.

\subsubsection{Simulations}

 Our simulation results for different parameters are compiled in the bottom row of Fig.~\ref{fig_Bvar}.
 In the reference case of a constant magnetic field with $\kappa_1=0$ (Fig.~\ref{fig_Bvar}a), the effective attraction due to activity decreases when $\kappa_0$ increases, while the density maxima eventually disappear as their location shifts away from the potential center, just like the effective potential minima in a polar geometry, cf.\ Fig.~\ref{fig_Bconst}b.

 Increasing instead the slope $\kappa_1$ while keeping $\kappa_0=0$ fixed (Fig.~\ref{fig_Bvar}b),
 the effect on the density distribution (or the effective potential) is qualitatively similar to the case of an increasing magnitude of a constant magnetic field,
 as long as $\kappa_1\lesssim1$ remains sufficiently small.
 In particular, there is always a symmetry with respect to $x=0$, since the magnitude of the Lorentz force increases in the same way on both branches.
When further increasing $\kappa_1$, another effect appears to take over and a new maximum evolves at the center.
This behavior is caused by the magnetic field reversing its direction, a scenario for which similar observations were made 
under stochastic resetting (but without a confining potential) \cite{abdoli2020reset,abdoli2022resetactive}.
As laid out in Ref.~\onlinecite{abdoli2020reset} for a passive system, 
the particle trajectories resemble a sequence of half circles directed along the symmetry axis of the system in the region where the magnetic field changes its sign.
This results in an effective dynamical trapping close to the origin of the coordinate $x$ perpendicular to this flux.
For an active system, the resulting localization is particularly strong in a small region around $x=0$ \cite{abdoli2022resetactive}.
Here, we find that this effect prevails even in the stationary state of a confined system.
Specifically, we observe a stronger localization with increasing $\kappa_1$, as
the variation of the magnetic field becomes both sharper and more pronounced.
The two outer maxima presumably appear because the nearby particles are drawn more and more to the center and not due to effective wall accumulation, which is suppressed under large magnetic fields.

When breaking the symmetry by choosing a fixed offset $\kappa_0=2$ (Fig.~\ref{fig_Bvar}c) of the magnetic field, we observe once more a localization in the region around $x/d=-\kappa_0/\kappa_1$ at which $\kappa(x)=0$, which is shifted towards the origin from the negative half space when increasing $\kappa_1$.
If this slope parameter is small, the corresponding density peak gradually emerges from the peak at the left wall, strongly suppressing the wall accumulation.
For very large $\kappa_1$, there eventually emerges an additional third peak at the left wall.
In general, it is always more likely to find a particle in the positive half space, which is further away from the trapping region and where absolute value of $\kappa(x)$ is larger. 

To better understand the detailed mechanism behind the effective trapping induced by the interplay of activity and the inversion of the Lorentz force, we additionally fix $\kappa_1=8$ and, instead, consider different persistence times $\tau$ (Fig.~\ref{fig_Bvar}d).
Indeed, for a small (but finite) $\taua$, we clearly observe three distinct density peaks, which fully merge for large $\taua$.
We suspect two main reasons for this behavior.
First, when increasing the persistence time $\taua$, the trajectories of the active particles start to resemble perfect circles (or half circles within the trapping region), such that
less particles can escape by randomly changing their velocity direction, i.e., the effective trapping becomes stronger.
Second, the  self-propulsion velocity $v_0=\sqrt{\mathfrak{d}D_\text{a}/\tau}$ decreases when increasing $\taua$, which further results in larger radii of the circular trajectories, i.e., the effective trapping region increases.
Only for very small $\tau$, the central peak can no longer be observed (not shown), as the activity starts to become irrelevant.
The strict limit $\tau\rightarrow 0$ corresponds to a passive system with diffusivity $D_\text{t}(1+\Da)$, where the magnetic field has no effect.

\subsubsection{Theory}

  Now we compare the corresponding theoretical predictions, compiled in the top row of Fig.~\ref{fig_Bvar}, to our simulations.
  In general, the theory does not always reproduce the correct overall behavior,
  while the location of the maxima (as compared in the insets) is, in most cases, predicted quite accurately.

  For a constant magnetic field (Fig.~\ref{fig_Bvar}a) the qualitative agreement between the two approaches is excellent.
  Specifically, the nonmonotonic dependence of the density maxima on $\kappa_0$ is reproduced with high accuracy.
 We also notice that, upon comparing densities rather than effective potentials, the agreement between theory and simulations for large $\kappa_0$ is even more reassuring in view of the prediction in Eq.~\eqref{eq_limit}.
Similar conclusions can be drawn for an inhomogeneous magnetic field (Fig.~\ref{fig_Bvar}b), but only when the slope $\kappa_1$ is sufficiently small.

The most striking feature of confined AOUPs in the inhomogeneous magnetic field $\kappa(x)$ considered here is the strong localization in the region of its sign change.
Indeed, the corresponding density peak is predicted by our theory at the precise location found in simulations and expected by our argument for $\kappa(x)=0$ (Fig.~\ref{fig_Bvar}c).
Also the tendency of the peak on the right to move toward the center with increasing $\kappa_1$ is well captured.
The only exception is the system in Fig.~\ref{fig_Bvar}b,  where the two locations of vanishing potential curvature and inverting magnetic field coincide.
Here, the maximum evolves from the intermediate minimum in the potential center, which the theory fails to predict due to its local nature.

Other theoretical predictions are at most qualitative.
Since it is already known that the theory tends to overestimate certain effects of the activity \cite{activePair}, we cannot expect that the peculiar interplay with the magnetic field is fully reproduced within exactly the same parameter regime.
To get a feeling for the resulting discrepancies, we change the persistence time $\tau$ of the active noise (Fig.~\ref{fig_Bvar}d).
The strong quantitative deviations at relatively small $\tau$ point to the high importance of the effects induced by the inhomogeneous Lorentz force.
Although we  see that the theory is also capable of predicting three distinct density peaks, 
this effect becomes only prominent at larger $\tau$ than in simulations.
In turn, the strong trapping found in simulations for large $\tau$
is not captured and the theory starts to break down locally (see the shaded region in Fig.~\ref{fig_Bvar}d), as the effective diffusivity becomes negative.
While our approach, and specifically the identification of an effective potential, is then, strictly speaking, ill defined (we refrained here from including an empirical correction, as described in  Sec.~\ref{sec_ELP2d}),
this result allows us to infer a strong non-Gaussian shape of the velocity distribution  in this region along the lines of Refs.~\onlinecite{caprini2022spatial1,caprini2022spatial2}.

\section{Conclusions \label{sec_conclusion}}

We have derived an effective FPE for AOUPs with conservative interactions in the presence of an additional Lorentz force, thus providing generalizations of both the Fox approach and the so-called ``best Fokker-Planck approximation'' (BFPA).
In general, the latter theory, despite its inclusion of additional terms, 
is shown not to provide a useful alternative to currently available effective equilibrium theories.
Combining ideas from the analytic resummation of the BFPA to an ordinary differential equation
with the expansion performed in Refs.~\onlinecite{martin2020,fodor2016} might, however, provide an interesting direction for future research. 

For a system of AOUPs subjected to a Lorentz force,
our methods used to derive the BFPA turn out particularly useful.
In this case, we have demonstrated that the Fox-type version of our resulting effective FPE 
readily allows to predict the effective interactions in the stationary state.
Specifically, our effective potentials rationalize that the Lorentz force counteracts the tendency of active particles to climb up potential barriers.
 With this intuition gained, 
 we can further arrive at the qualitative conclusion that the presence of a Lorentz force suppresses both motility-induced phase separation \cite{buttinoni2013,stenhammar2014,cates_tailleur2014}
 and wall accumulation or wetting \cite{pototsky2012,yang2014,ni2015,neta2021,wittmannbrader2016}.
 Moreover, the escape over a potential barrier \cite{sharma2017} is expected to be aggravated by a constant magnetic field.
 As a further step, the effective potentials provide simple insights in the response of an active system to
 an inhomogeneous magnetic field.
 In our system, we observe up to three distinct density maxima,
 as the region where the magnetic field reverses its direction acts as an effective trap. 
 This localization gets stronger when increasing either the gradient of the magnetic field or the persistence time of the active particles.

Beyond the stationary properties investigated here, the coupling of activity and magnetic field gives rise to rich nonequilibrium phenomena such as enhanced dynamics \cite{vuijk2020magnetic}, boundary currents \cite{abdoli2020reset} and directed transport \cite{muzzeddu2022active}, to name a few.
It will be interesting to see in how far these dynamical effects can be extracted from our effective treatment.
Another perspective is to compare our models and results to 
the dynamics of chiral active particles \cite{vanteeffelen2008chiral,sevilla2016chiral, caprini2019cAOUPs}, 
specifically in confinement \cite{caprini2023chiral}. 
While it has been shown for the active Brownian models that the same odd-diffusive behavior is observed on the single-particle level \cite{muzzeddu2022active}, it is not clear whether the resulting FPEs are also equivalent in systems of interacting particles, as one also needs to account for the mobility. 
Moreover, the chiral AOUP model in Ref.~\onlinecite{caprini2019cAOUPs} does not admit a workable effective equilibrium description of single particles, like the one derived here for AOUPs under Lorentz force,  even for simple confining potentials.

The overdamped limit considered here corresponds to the description of dynamics on a time scale greater than the velocity correlation time $m/\gamma$, where $m$ is the mass of the particle.
In equilibrium systems with Lorentz force, this corresponds to the small-mass limit, which is distinctive from the large-friction limit \cite{chun2018}. 
In the small-mass limit, the diffusion tensor, with its antisymmetric part, still captures the curving effect of the Lorentz force on the particles' trajectory.
Our present theory for active particles inherently assumes that $\tau$, the persistence time of the active fluctuations, is greater than $m/\gamma$, such that the limit $\tau\rightarrow0$ in Eq.~\eqref{eq_GammaFmag} still carries the footprint of the coupling between activity and Lorentz force and is thus distinct from a passive particle in a magnetic field.
We leave it future work to explore a scenario with these limits being reversed.

 As demonstrated by the example of the calculations performed in this paper,
the early works on colored noise can still be relevant in the contemporary context of active matter.
Specifically, a topic of rapidly growing interest is inertial active motion \cite{scholz2018,breoni2020,sandoval2020,caprini2021velocityCORRinertia,sprenger2021,loewenREVinertia2020},
also featuring according variants of the AOUP model \cite{nguyen2021,caprini2021inertiaAOUP,sahoo22,sahoo23,sahoo24,sprenger2023,caprini2023entropons,caprini2024tapping}.
A logical strategy to derive effective probability distributions for such systems 
would be to build upon existing versions of effective underdamped FPEs  \cite{sancho1980,hernandezmachado1984,lindenberg1983MC,lindenberg1984MC,fronzoni1986inertialFOX,sancho1988inertialFOX,hanggi1989inertialUCNA}.
This will be particularly interesting for systems subject to a Lorentz force.
In this case, it should be explored in how far the previously discussed noninterchangability of overdamped and white-noise limits also affects the resulting  effective equilibrium descriptions.
 Another notable aspect of the old literature is that the main focus usually lies on the more general situation involving multiplicative colored noise. 
The relevance of a multiplicative noise in the context of activity has already been recognized for describing a spatial modulation of the self-propulsion velocity in the AOUP model~\cite{caprini2022spatial1,caprini2022spatial2}, 
which also opens a new avenue to study spatially inhomogeneous systems subjected to a Lorentz force.

\acknowledgments
The authors would like to thank Lorenzo Caprini and Hartmut L\"owen for stimulating discussions and critical reading of the manuscript.
Funding by the Deutsche Forschungsgemeinschaft (DFG)  through the SPP 2265 under grant numbers LO 418/25-2 (IA) and WI 5527/1-2 (RW) is gratefully acknowledged.

\appendix

\section{Description of numerical simulations \label{app_numerics}}
To validate our theoretical predictions, we perform Brownian dynamics simulations using the Langevin equations of motion. In the presence of an external magnetic field, the overdamped limit of the Langevin equation requires a proper small-mass limit to avoid unphysical values for velocity-dependent variables such as fluxes~\cite{vuijk2019magnetic}. 
Therefore, we numerically integrate the underdamped Langevin equations, generalizing Eq.~\eqref{eq_AOUPs} to 
\begin{equation}
m \ddot{x}_\alpha(t) = -\gamma\dot{x}_\alpha(t) + \tilde{F}_\alpha(\{x_\beta\}) + I_\text{t}\gamma\xi_\alpha(t) + \gamma\chi_\alpha(t)\,.
\label{eq_AOUPs_un}
\end{equation}
To mimic the small-mass limit, we choose the mass $m$ as $m=0.02 \gamma d^2/D_\text{t}$  and the integration time step $\Delta t$ is chosen as $\Delta t=10^{-5} d^2/D_\text{t}$. 
Here, $\tilde{F}_\alpha:={F}_\alpha+{F}^\text{L}_\alpha$ where $F_\alpha$ is the conservative forces and ${F}^\text{L}_\alpha=q\epsilon_{\alpha\mu\nu}\dot{x}_\mu B_\nu$ is Lorentz forces with $q$, $B_\nu$, and $\epsilon_{\alpha\mu\nu}$ being the charge, the external magnetic field, and the Levi-Civita symbol, respectively.  The stationary stochastic processes $\chi_\alpha(t)$ in \eqref{eq_AOUPs_un} evolve in time according to
\begin{equation}
\dot{\chi}_\alpha(t)=-\frac{\chi_\alpha(t)}{\taua}+\frac{\eta_\alpha(t)}{\taua}\,,
\label{eq_AOUPsDEF_un}
\end{equation}
where $\xi_\alpha$ and $\eta_\alpha$ are Gaussian white noises with zero means and $\langle\xi_\alpha(t)\xi_\beta(t')\rangle\!=\!2D_\text{t}\delta_{\alpha\beta}\delta(t-t')$ and
$\langle\eta_\alpha(t)\eta_\beta(t')\rangle\!=\!2D_\text{a}\delta_{\alpha\beta}\delta(t-t')$,
where $D_\text{t}$ and $D_\text{a}$ are the diffusion coefficients characterizing passive Brownian motion and the active propulsion, respectively. As in the main text, $I_\text{t}\in\{0,1\}$ indicates whether or not the (translational) Brownian noise $\xi_\alpha(t)$ is present and $\taua$ is the persistence time of the active motion.

\section{Derivation of the multicomponent BFPA with Lorentz force \label{app_derivation}}
 
In this appendix, we detail the derivation of our central results discussed in the main text, 
the multicomponent BFPA, Eq.~\eqref{eq_GammaB}, and the Fox approximation, Eq.~\eqref{eq_GammaFmag}, for charged AOUPs in a magnetic field.
Both cases can be obtained from the more general result, the BFPA in presence of a magnetic field.
 To this end, we introduce for later notational convenience a general short notation 
\begin{equation}
 A_{\alpha\beta}^G:=A_{\alpha\gamma}G_{\gamma\beta}\,,\ \ \ {}^G\!\!A_{\alpha\beta}:=G_{\alpha\gamma}A_{\gamma\beta}
\end{equation}
denoting the contraction between the magnetic tensor $G_{\alpha\beta}(\{x_\gamma\})$ from Eq.~\eqref{eq_Gmag} and an arbitrary tensor $A_{\alpha\beta}$ (or analogously a vector).
Due to the asymmetric nature of $G_{\alpha\beta}$ the order of indices is important here.
The derivation in the absence of a magnetic field then simply follows from setting $A_{\alpha\beta}^G\rightarrow A_{\alpha\beta}$ and ${}^G\!\!A_{\alpha\beta}\rightarrow A_{\alpha\beta}$.

Our starting point is the joint Fokker-Planck equation from Eq.~\eqref{eq_FPEjoint} of the main text. 
As for all approximation schemes starting from a formally exact FPE, the contribution of thermal noise in Eq.~\eqref{eq_FPEjoint} will eventually only constitute a trivial additive factor
to the effective diffusion tensor as in Eq.~\eqref{eq_GammaF}. 
Thus, we set $I_\text{t}=0$ for simplicity of the following presentation,
since it can be simply reintroduced afterwards, as discussed in the context of Fox's approach \cite{sharma2017}. 
Introducing the operators
\begin{align}
\mathcal{L}_a=
-D_\text{t}\partial_\beta^G\beta F_\beta(\{x_\gamma\})\equiv-D_\text{t}\partial_\alpha\beta\,{}^G\!\!F_\alpha(\{x_\gamma\})\,, \label{eq_Laapp}\end{align}
which only depends on the spatial variables,
\begin{align}
\mathcal{L}_b=\frac{1}{\taua}\frac{\partial}{\partial\chi_\alpha}\chi_\alpha+\frac{D_\text{a}}{\taua^2}\frac{\partial}{\partial\chi_\alpha}\frac{\partial}{\partial\chi_\alpha}\,, \label{eq_Lbapp}\end{align}
which only depends on the values of the Ornstein-Uhlenbeck processes, and 
\begin{align}
\mathcal{L}_1=
-\chi_\beta \partial_\beta^G\,,\label{eq_L1app}
\end{align}
containing cross terms,
we reexpress Eq.~\eqref{eq_FPEjoint} as
\begin{equation}
\partial_tP_N(\{x_\alpha\},\{\chi_\alpha\},t)=(\mathcal{L}_a+\mathcal{L}_b+\mathcal{L}_1)P_N(\{x_\alpha\},\{\chi_\alpha\},t)\,,
\label{eq_FPEjointLiapp}
\end{equation}
 just as in Eq.~\eqref{eq_FPEjointLi}.

\subsection{Projecting onto equilibrium processes \label{sec_PO}}

 As a first step, we sketch the derivation of Eq.~\eqref{eq_FPEformal} from Eq.~\eqref{eq_FPEjointLiapp}, which does not require detailed knowledge of the operators $\mathcal{L}_a$, $\mathcal{L}_b$ and $\mathcal{L}_1$.
Here, we use the projection-operator formalism introduced by Zwanzig~\cite{zwanzigPO}
and employed in the present context of colored noise by Grigolini and coworkers \cite{faetti1988,grigolini1986,grigolini1985ACP1}.

Let us further consider the mixed operator $\mathcal{L}_1$ as a perturbation
to the Markovian operator $\mathcal{L}_0:=\mathcal{L}_a+\mathcal{L}_b$. 
This initial separation explicitly requires the assumption of small fluctuations, i.e., a small persistence time $\taua$ in Eq.~\eqref{eq_AOUPsCORR}.
Then we switch to the interaction picture with
\begin{align}
\tilde{P}_N(t)&:=\exp(-\mathcal{L}_0t)P_N(t)\,,\label{eq_tildePN}\\
\tilde{\mathcal{L}}_1(t)&:=\exp(-\mathcal{L}_0t)\mathcal{L}_1\exp(\mathcal{L}_0t)
\end{align}
such that Eq.~\eqref{eq_FPEjointLi} can be rewritten as
\begin{equation}
\partial_t\tilde{P}_N(t)=\tilde{\mathcal{L}}_1(t)\tilde{P}_N(t)\,.
\label{eq_FPEjointLiINT}
\end{equation}

Now we introduce the generalized projection operator $\mathcal{P}$,
defined through its action on an observable $O$ as
\begin{align}
\mathcal{P}O
:=R_N(\{\chi_\alpha\})\int\upd\boldsymbol{\chi}^NO
\end{align}
onto the equilibrium distribution, Eq.~\eqref{eq_RN}, of the AOUPs, such that $\mathcal{P}P_N=f_NR_N$, and define
\begin{align}
\tilde{P}^{(1)}_N(t)&:=\mathcal{P}\tilde{P}_N(t)\,\label{eq_tildePN1}\\
\tilde{P}^{(2)}_N(t)&:=(1-\mathcal{P})\tilde{P}_N(t)\,,
\end{align}
such that $\tilde{P}_N=\tilde{P}^{(1)}_N+\tilde{P}^{(2)}_N$.
Then we rewrite Eq.~\eqref{eq_FPEjointLiINT} in the form of the two coupled equations
\begin{align}
\partial_t\tilde{P}^{(1)}_N(t)&=\mathcal{P}\tilde{\mathcal{L}}_1(t)\left(\tilde{P}^{(1)}_N(t)+\tilde{P}^{(2)}_N(t)\right)\,,\label{eq_FPEjointLiINTx1}\\
\partial_t\tilde{P}^{(2)}_N(t)&=(1-\mathcal{P})\tilde{\mathcal{L}}_1(t)\left(\tilde{P}^{(1)}_N(t)+\tilde{P}^{(2)}_N(t)\right)\,
\label{eq_FPEjointLiINTx2}
\end{align}
and formally solve Eq.~\eqref{eq_FPEjointLiINTx2}, which gives~\cite{grigolini1985ACP1}
\begin{align}
\tilde{P}^{(2)}_N(t)=&\ \overleftarrow{\exp}\left(\int_0^t\upd s(1-\mathcal{P})\tilde{\mathcal{L}}_1(s)\right)\tilde{P}^{(2)}_N(0)\cr
&+\int_0^t\upd s\;\overleftarrow{\exp}\left(\int_s^t\upd s'(1-\mathcal{P})\tilde{\mathcal{L}}_1(s')\right)\cr
&\ \ \ \ \ \ \ \times(1-\mathcal{P})\tilde{\mathcal{L}}_1(s)\tilde{P}^{(1)}_N(s)
\label{eq_FPEjointLiINTx3}
\end{align}
where the arrow indicates that we are dealing with a time-ordered exponential, since we use the interaction picture.

Before substituting Eq.~\eqref{eq_FPEjointLiINTx3} into Eq.~\eqref{eq_FPEjointLiINTx1}
we make the following simplifications:
Firstly, we assume that the first ``preparation'' term in Eq.~\eqref{eq_FPEjointLiINTx3} vanishes,
which depends on the initial conditions $\tilde{P}^{(2)}_N(0)$, and thus is irrelevant for large observation times, $t\rightarrow\infty$~\cite{grigolini1985ACP1}.
Secondly, we restrict ourselves to second order in the perturbation term $\tilde{\mathcal{L}}_1(t)$,
which is equivalent to setting the time-ordered exponential to one.
Finally, we drop the term linear in $\tilde{\mathcal{L}}_1(t)$, i.e., the first term in Eq.~\eqref{eq_FPEjointLiINTx1},
which eventually drops out of the calculation since $\langle\chi_\alpha(t)\rangle=0$.
The result is
\begin{align}
\partial_t\tilde{P}^{(1)}_N(t)=\int_0^t\upd s\, \mathcal{P}\tilde{\mathcal{L}}_1(t)  (1-\mathcal{P})\tilde{\mathcal{L}}_1(s)\tilde{P}^{(1)}_N(s)\,.
\label{eq_FPEproject}
\end{align}

To turn Eq.~\eqref{eq_FPEproject} into an equation for
\begin{equation}
 f_N(\{x_\alpha\},t)=R_N(\{\chi_\alpha\})^{-1}\mathcal{P}P_N(\{x_\alpha\},\{\chi_\alpha\},t)
\end{equation}
we make use of the commutation rules 
\begin{align}
[\mathcal{P},\mathcal{L}_a]=[\mathcal{P},e^{\mathcal{L}_at}]=[\mathcal{P},\partial_t]=0\label{eq_commutation}
\end{align}
 and the properties
\begin{align}
\mathcal{P}e^{\mathcal{L}_bt}=e^{\mathcal{L}_bt}\mathcal{P}=\mathcal{P}\,
\end{align}
of a projection $\mathcal{P}=\mathcal{P}^2$ onto the equilibrium distribution corresponding to $\mathcal{L}_b$.
As we can write $\mathcal{P}=\prod_\alpha p_{\chi_\alpha}$ with
$p_\chi A(\chi,t):=\rho(\chi)\int\upd\chi A(\chi,t)$,
these relation are the same as for one component~\cite{grigolini1985ACP2}.
Then we find with the help of Eqs.~\eqref{eq_tildePN} and~\eqref{eq_tildePN1}
\begin{align}
\partial_tf_N(t)&=R_N^{-1}\mathcal{P}\partial_tP_N(t)=R_N^{-1}\mathcal{P}\partial_t\,e^{\mathcal{L}_at+\mathcal{L}_bt}\tilde{P}_N(t)\nonumber\\
&=R_N^{-1}\partial_t\,e^{\mathcal{L}_at}\mathcal{P}\tilde{P}_N(t)\cr
&=R_N^{-1} \left(\mathcal{L}_a R_Nf_N(t)+ e^{\mathcal{L}_at} \partial_t\tilde{P}^{(1)}_N(t)\right).
\end{align}
Now we substitute Eq.~\eqref{eq_FPEproject} into the second term,
which gives the formal result
\begin{equation}
\partial_tf_N(t)=\mathcal{L}_af_N(t)+\int_0^t\upd s\,K(t-s) f_N(s)
\label{eq_FPEformalB}
\end{equation}
with the convolution kernel
\begin{align}
K(t-s)&=R_N^{-1} e^{\mathcal{L}_at}\mathcal{P}\tilde{\mathcal{L}}_1(t)  (1-\mathcal{P})\tilde{\mathcal{L}}_1(s)e^{-\mathcal{L}_as}\mathcal{P}R_N\nonumber\\
&=R_N^{-1} \mathcal{P}\mathcal{L}_1e^{\mathcal{L}_0t} (1-\mathcal{P})e^{-\mathcal{L}_0s}\mathcal{L}_1\mathcal{P}R_N
\label{eq_kernalALLG}
\end{align}
where similar modifications have been employed.

At second perturbation order in $\mathcal{L}_1$ we may neglect the history dependence  
expressed by the convolution form ($K(t-s) f_N(s)\simeq K(s) f_N(t-s)$) of the second term in Eq.~\eqref{eq_FPEformalB} and set \cite{faetti1988,grigolini1985ACP2}
\begin{equation}
K(s) f_N(t-s)\simeq K(s) e^{-\mathcal{L}_as}f_N(t)\,
\label{eq_DBappBFPA}
\end{equation}
within the integrand.
 This is where we enforce detailed balance.
Since the stochastic variables $\chi_\alpha$ are associated with AOUPs,
we can write the kernel in Eq.~\eqref{eq_kernalALLG} as 
\begin{align}
K(s'\!=t-s)&=R_N^{-1} \mathcal{P}\chi_\alpha e^{\mathcal{L}_bt} (1-\mathcal{P})e^{-\mathcal{L}_bs}\chi_\beta\mathcal{P}R_N\cr
&\ \ \ \times\partial^G_\alpha e^{\mathcal{L}_a(t-s)}\partial^G_\beta\cr
&= R_N^{-1} \mathcal{P}\chi_\alpha e^{\mathcal{L}_bs'}\chi_\beta R_N \partial^G_\alpha e^{\mathcal{L}_as'}\partial^G_\beta\cr
&=\langle\chi_\alpha(0)\chi_\beta(s')\rangle\partial^G_\alpha e^{\mathcal{L}_as'}\partial^G_\beta
\label{eq_kernalSPEC}
\end{align}
where we inserted Eq.~\eqref{eq_L1app} and used the averages
\begin{align}
\mathcal{P}\chi_\alpha R_N&=\langle\chi_\alpha\rangle=0\,,\cr
  \mathcal{P}\chi_\alpha e^{\mathcal{L}_bs'}\chi_\beta R_N&=R_N\langle\chi_\alpha(0)\chi_\beta(s')\rangle\,. 
    \end{align}
Note that in the second line $\exp(\mathcal{L}_bs')$ denotes a translation in time, which does not act on the equilibrium distribution $R_N$.
At this stage it is now also obvious that it is justified to drop the term linear in $\mathcal{L}_1$ before writing down Eq.~\eqref{eq_FPEproject}.
Combining Eqs.~\eqref{eq_FPEformalB},~\eqref{eq_DBappBFPA} and~\eqref{eq_kernalSPEC}, we obtain
\begin{equation}
\partial_tf_N(t)=
\mathcal{L}_af_N(t)+\int_0^t\upd s\left\langle \mathcal{L}_\chi(0)\boldsymbol{\mathcal{L}}(s)\right\rangle f_N(t)\!\!\!
\label{eq_FPEformalAPP}
\end{equation}
with
 \begin{align}
  \boldsymbol{\mathcal{L}}(s):=e^{\mathcal{L}_as}\mathcal{L}_\chi(s)e^{-\mathcal{L}_as} \label{eq_FPEformalLAST}
 \end{align}
 and 
 \begin{align}
\mathcal{L}_\chi(t)=-\chi_\beta(t) \partial_\beta^G\,.\label{eq_Lchiapp}
\end{align}
We have thus obtained Eq.~\eqref{eq_FPEformal} for generalized operators also including interactions with a magnetic field.

\subsection{
Identifying the diffusion tensor \label{sec_BFPAmulti}}

The second step is to bring Eq.~\eqref{eq_FPEformalAPP} to the form of Eq.~\eqref{eq_FPEgenLORENTZ}.
We immediately notice the equivalence of the first term, cf.\ Eq.~\eqref{eq_Laapp}.
 The integral 
\begin{equation}
\partial_\alpha^G\partial_\beta\mathcal{D}^t_{\alpha\beta}(t):=
D_\text{t}^{-1}\int_0^t\upd s\left\langle \mathcal{L}_\chi(0)\boldsymbol{\mathcal{L}}(s)\right\rangle
\label{eq_FPEformalAPPint}
\end{equation}
in the second term must therefore be recast in the form of a second spatial derivative of a diffusion tensor $\mathcal{D}^t_{\alpha\beta}(\{x_\gamma\},t)$,
which, in general, apparently depends on time (supersctipt $t$).
To carry out the integration, we follow the strategy from Ref.~\onlinecite{lindenberg1984MC} and process a hierarchy of commutators.
 To this end we first expand the last three operators appearing in Eq.~\eqref{eq_FPEformal}, cf.\ Eq.~\eqref{eq_FPEformalLAST}, into a Taylor series, which yields
 \begin{align}
\boldsymbol{\mathcal{L}}(s)=&\mathcal{L}_\chi(s)+s[\mathcal{L}_a,\mathcal{L}_\chi(s)]+\frac{s^2}{2!}[\mathcal{L}_a,[\mathcal{L}_a,\mathcal{L}_\chi(s)]]\cr
&+\frac{s^3}{3!}[\mathcal{L}_a,[\mathcal{L}_a,[\mathcal{L}_a,\mathcal{L}_\chi(s)]]]+\ldots\,.
\end{align}
  At the $n$th order of such an expansion we define the commutator 
\begin{align}
C^{(n)}(s):=\frac{s^n}{n!}[\mathcal{L}_a,[\mathcal{L}_a,\ldots[\mathcal{L}_a,\mathcal{L}_\chi(s)]]]
\label{eq_CnDEF}
\end{align}
containing the operator $\mathcal{L}_a$ $n$ times,
so that we can write
\begin{align}
\boldsymbol{\mathcal{L}}(s)=\sum_{n=0}^\infty C^{(n)}(s)=-\chi_\beta(s)\,\partial^G_\alpha\sum_{n=0}^\infty\frac{s^n}{n!}\Delta^{(n)}_{\alpha\beta}
\label{eq_LLs}
\end{align}
introducing the auxiliary tensor $\Delta^{(n)}_{\alpha\beta}$ in the last step,
which is implicitly defined as the generator of the commutator 
\begin{align}
\!\!\!C^{(n)}(s)=-\chi_\beta(s)\,\partial^G_\alpha\frac{s^n}{n!}\Delta^{(n)}_{\alpha\beta}\equiv-\chi_\beta(s)\,\partial_\alpha\frac{s^n}{n!}{}^G\!\!\Delta^{(n)}_{\alpha\beta}\,.
\label{eq_Cn}
\end{align}
Its tensorial nature comes from the different possible combinations of the components of $\partial^G_\alpha$ and $\chi_\beta(s)$ after placing the latter in front of the commutator in Eq.~\eqref{eq_CnDEF}.

We can determine a closed expression for $\Delta^{(n)}_{\alpha\beta}$ as follows.
The term of zero order is given by
\begin{align}
\Delta^{(0)}_{\alpha\beta}=\delta_{\alpha\beta}
\label{eq_Delta0}
\end{align}
since $C^{(0)}(s)=\mathcal{L}_\chi(s)=-\chi_\alpha(s)\partial^G_\alpha$.
If we further assume that $C^{(n-1)}(s)$ is known, we can rewrite Eq.~\eqref{eq_CnDEF} as
\begin{align}
C^{(n)}(s)&=\frac{s}{n}[\mathcal{L}_a,C^{(n-1)}(s)]\cr&=D_\text{t}\chi_\beta(s)\frac{s^n}{n!}\left[\partial^G_\gamma\beta F_\gamma,\partial^G_\alpha\Delta^{(n-1)}_{\alpha\beta}\right],
\label{eq_CnCOM}
\end{align}
making use of Eq.~\eqref{eq_Cn} at order $n-1$ in the last step.
Comparing now to Eq.~\eqref{eq_Cn} at order $n$ we can identify the commutator in the last expression of Eq.~\eqref{eq_CnCOM} with $-D_\text{t}^{-1}\partial^G_\alpha\Delta^{(n)}_{\alpha\beta}$ and thus deduce the 
recursive relation
\begin{align}\!\!\!
\partial^G_\alpha\Delta^{(n)}_{\alpha\beta}=D_\text{t}\left(\partial^G_\alpha\Delta^{(n-1)}_{\alpha\beta} \partial^G_\gamma\beta F_\gamma -\partial^G_\gamma\beta F_\gamma\partial^G_\alpha\Delta^{(n-1)}_{\alpha\beta}\right).\!\!\!
\label{eq_Deltan23}
\end{align}
This means that all terms in Eq.~\eqref{eq_LLs} are formally known.
Finally, we are able to proof in the next paragraph that the auxiliary tensor $\Delta^{(n)}_{\alpha\beta}$ (multiplied from the left with the magnetic tensor) it is recursively given by 
\begin{align}
{}^G\!\!\Delta^{(n)}_{\alpha\beta}= D_\text{t}\beta\left({}^G\!\!\Delta^{(n-1)}_{\gamma\beta} (\partial_\gamma {}^G\!\!F_\alpha) - {}^G\!\!F_\gamma(\partial_\gamma{}^G\!\!\Delta^{(n-1)}_{\alpha\beta})\right)
\label{eq_Deltan23b}
\end{align}
and the zero order, $n=0$, in Eq.~\eqref{eq_Delta0}.
This closed form implies that
\begin{align}
 {}^G\!\!\Delta^{(n)}_{\alpha\beta}f_N=f_N{}^G\!\!\Delta^{(n)}_{\alpha\beta}\,,
 \label{eq_noop}
\end{align}
i.e., $\Delta^{(n)}_{\alpha\beta}$ is not an operator.

To proof Eq.~\eqref{eq_Deltan23b} we show by induction that the relation
\begin{align}
\partial_\alpha{}^G\!\!\Delta^{(n)}_{\alpha\beta}f_N=f_N(\partial_\alpha{}^G\!\!\Delta^{(n)}_{\alpha\beta})+(\partial_\alpha f_N){}^G\!\!\Delta^{(n)}_{\alpha\beta}\,,
\label{eq_dDcomm}
\end{align}
  which is necessary if we require Eq.~\eqref{eq_noop} to hold, is fulfilled at any order $n$ for an arbitrary distribution function $f_N$.
With the help of Eq.~\eqref{eq_Deltan23}, we can explicitly calculate the derivative on the left-hand side of Eq.~\eqref{eq_dDcomm}.
Arranging the terms by the order of the derivative of $f_N$, we obtain
\begin{align}
\partial_\alpha{}^G\!\!\Delta^{(n)}_{\alpha\beta}f_N=f_N(\partial_\alpha{}^G\!\!\Delta^{(n)}_{\alpha\beta})+X_{\beta}+(\partial_\alpha\partial_\gamma f_N)\,Y_{\alpha\beta\gamma}\,.
\label{eq_dDgen}
\end{align}
where the tensor
\begin{align}
 \frac{Y_{\alpha\beta\gamma}}{D_\text{t}\beta}={}^G\!\!\Delta^{(n-1)}_{\alpha\beta}{}^G\!\! F_\gamma - {}^G\!\!F_\gamma {}^G\!\!\Delta^{(n-1)}_{\alpha\beta}=0
\end{align}
vanishes by the induction assumption that ${}^G\!\!\Delta^{(n-1)}_{\alpha\beta}$ given in the form of Eq.~\eqref{eq_Deltan23b} does not operate on $F_\gamma$ or $f_N$.
The vector $X_\beta$ containing the first derivatives of $f_N$ reads
\begin{align}
\!\!\!\!\!\!\frac{X_\beta}{D_\text{t}\beta}&= 
(\partial_\alpha f_N){}^G\!\!\Delta^{(n-1)}_{\alpha\beta} \partial_\gamma {}^G\!\!F_\gamma
+(\partial_\gamma f_N)\partial_\alpha{}^G\!\!\Delta^{(n-1)}_{\alpha\beta} {}^G\!\!F_\gamma\cr
&\ \ \ \ \, -(\partial_\gamma f_N) {}^G\!\!F_\gamma\partial_\alpha{}^G\!\!\Delta^{(n-1)}_{\alpha\beta}
-(\partial_\alpha f_N)\partial_\gamma {}^G\!\!F_\gamma{}^G\!\!\Delta^{(n-1)}_{\alpha\beta}\cr
&=(\partial_\gamma f_N){}^G\!\!\Delta^{(n-1)}_{\alpha\beta} \partial_\alpha {}^G\!\!F_\gamma-(\partial_\alpha f_N){}^G\!\!F_\gamma \partial_\gamma {}^G\!\!\Delta^{(n-1)}_{\alpha\beta}\!\!,\!\!\!\!\!\!
\end{align}
where the last line again results from the induction assumption.
After exchanging the summation indices $\alpha$ and $\gamma$ in the first term, we find that $X_\beta=(\partial_\alpha f_N)\Delta^{(n)}_{\alpha\beta}$ if $\Delta^{(n)}_{\alpha\beta}$ is given by Eq.~\eqref{eq_Deltan23b}
and have thus established the equivalence between Eq.~\eqref{eq_dDcomm} and Eq.~\eqref{eq_dDgen}.
This proofs that $\Delta^{(n)}_{\alpha\beta}$, written as in Eq.~\eqref{eq_Deltan23b}, is an ordinary tensor for any order $n$.

Returning to Eq.~\eqref{eq_FPEformalAPPint}, we substitute Eqs.~\eqref{eq_Lchiapp} and~\eqref{eq_LLs} and obtain the integral
\begin{align}
\!\partial_\beta^G\partial_\gamma\mathcal{D}^t_{\gamma\beta}(t)&= D_\text{t}^{-1}\!\!\int_0^t\upd s\left\langle \chi_\alpha(0)\partial_\alpha^G  \chi_\beta(s)\partial_\gamma^G\sum_{n=0}^\infty\frac{s^n}{n!}\Delta^{(n)}_{\gamma\beta}\right\rangle  \nonumber\\
&=\int_0^t\upd s \frac{\Da}{\taua}e^{-\frac{s}{\taua}}\partial_\beta^G \partial_\gamma\sum_{n=0}^\infty\frac{s^n}{n!}{}^G\!\!\Delta^{(n)}_{\gamma\beta}\,,\!\!\!\!\!\!\!\!\!\!\!\!\!\!\!\!\!\!\!\!\!\!\!\!\!\!\!\!
\label{eq_LLi}
\end{align}
where we used Eq.~\eqref{eq_AOUPsCORR} to rewrite the correlator $\left\langle \chi_\alpha(0)\chi_\beta(s)\right\rangle$ in the second line.
We thus identify the dimensionless effective diffusion tensor
\begin{align}
\mathcal{D}^t_{\alpha\beta}(\{x_\gamma\},t)=\frac{\Da}{\taua}\int_0^t\upd s \,e^{-\frac{s}{\taua}}\sum_{n=0}^\infty\frac{s^n}{n!}{}^G\!\!\Delta^{(n)}_{\alpha\beta}
\label{eq_DeffALLGt}
\end{align}
in its general time-dependent form.
 In the absence of a magnetic field, the above expression reduces to the result established in Ref.~\onlinecite{lindenberg1984MC}.

\subsection{
Approximations for the diffusion tensor
\label{sec_BFPAmulti2}}

As a final step we must solve the integral in Eq.~\eqref{eq_DeffALLGt}
to obtain a workable expression for the diffusion tensor entering in Eq.~\eqref{eq_FPEgenLORENTZ}.
To this end, we proceed as in the earlier derivations of the one-component BFPA \cite{grigolini1986,masoliver1987} and make the standard 
assumption that the observation time $t$ is large compared to the correlation time $\taua$.
Hence, we replace the upper integration limit in Eq.~\eqref{eq_FPEproject} with infinity and find
the time-independent diffusion tensor 
\begin{align}
\mathcal{D}_{\alpha\beta}(\{x_\gamma\})=\lim_{t\rightarrow\infty}\mathcal{D}^t_{\alpha\beta}(\{x_\gamma\},t)
\end{align}
which enters in Eq.~\eqref{eq_FPEgenLORENTZ}.

This approximation allows for an explicit calculation of $\mathcal{D}_{\alpha\beta}$, since
the integral in Eq.~\eqref{eq_DeffALLGt} becomes a Laplace transform. 
Recalling the standard integral
 \begin{align}
\int_0^\infty\upd s \,e^{-as}s^n=\frac{n!}{a^{n+1}}\,
\end{align}
we obtain
\begin{align}
\mathcal{D}_{\alpha\beta}=\Da \sum_{n=0}^\infty \tilde{\mathcal{D}}^{(n)}_{\alpha\beta} =\Da \sum_{n=0}^\infty \taua^{n}\,{}^G\!\!\Delta^{(n)}_{\alpha\beta} \,,
\label{eq_DeffALLG}
\end{align}
where we defined 
 \begin{align}
  \tilde{\mathcal{D}}^{(n)}_{\alpha\beta}:= \taua^{n}{}^G\!\!\Delta^{(n)}_{\alpha\beta}\,.
  \label{eq_DeffnALLG}
 \end{align}
Substituting Eq.~\eqref{eq_DeffnALLG} into Eq.~\eqref{eq_Deltan23b} yields
\begin{align}\!\!
\tilde{\mathcal{D}}^{(n)}_{\alpha\beta}&=\taua D_\text{t}\beta\left(\tilde{\mathcal{D}}^{(n-1)}_{\gamma\beta} \partial_\gamma {}^G\!\!F_\alpha - {}^G\!\!F_\gamma\partial_\gamma\tilde{\mathcal{D}}^{(n-1)}_{\alpha\beta}\right),
\label{eq_Dn23b}
\end{align}
for $n\geq1$, while the initial value 
\begin{align}
 \tilde{\mathcal{D}}^{(0)}_{\alpha\beta}={}^G\!\delta_{\alpha\beta}=G_{\alpha\beta}
\end{align}
 follows directly from inserting Eq.~\eqref{eq_Delta0} into Eq.~\eqref{eq_DeffnALLG} with $n=0$.

At this stage, it is already formally  possible to determine the effective diffusion tensor $\mathcal{D}_{\alpha\beta}$
by resummation according to Eq.~\eqref{eq_DeffALLG} of the coefficients $\tilde{\mathcal{D}}^{(n)}_{\alpha\beta}$, 
given by Eq.~\eqref{eq_Dn23b} for the general case or by Eq.~\eqref{eq_DeffALLGsum} of the main text for the case without a magnetic field.
To obtain a more compact representation of $\mathcal{D}_{\alpha\beta}$, we write
\begin{align}
 \sum_{n=0}^\infty \tilde{\mathcal{D}}^{(n)}_{\alpha\beta}=G_{\alpha\beta}+\sum_{n=1}^\infty \tilde{\mathcal{D}}^{(n)}_{\alpha\beta}\,
\end{align}
and substitute Eq.~\eqref{eq_Dn23b} into the expression on the right-hand side.
With the definition in Eq.~\eqref{eq_DeffALLG} we then find the differential equation
\begin{align}\!\!
\mathcal{D}_{\alpha\beta}&=\Da G_{\alpha\beta}+\taua D_\text{t}\beta\left(\mathcal{D}_{\gamma\beta} \partial_\gamma {}^G\!\!F_\alpha - {}^G\!\!F_\gamma\partial_\gamma\mathcal{D}_{\alpha\beta}\right).
\label{eq_D23b}
\end{align}
The solution of this equation is formally equivalent to the infinite sum over $\tilde{\mathcal{D}}^{(n)}_{\alpha\beta}$
but generally not available in a closed form.
The representation in Eq.~\eqref{eq_DeffALLG}, on the other hand,
constitutes a power series in the persistence time $\taua$ which might not converge or only converge slowly.

An alternative expansion for $\mathcal{D}_{\alpha\beta}$ can be found by iteration of Eq.~\eqref{eq_D23b},
which allows us to establish a connection with the multicomponent Fox result \cite{grigolini1986}.
To provide a more compact notation we first rewrite Eq.~\eqref{eq_D23b} as
\begin{align}\!\!
\!\mathcal{D}_{\gamma\beta}\,(\delta_{\alpha\gamma}-\taua D_\text{t}\beta\,\partial_\gamma {}^G\!\!F_\alpha)=\Da G_{\alpha\beta}-\taua D_\text{t} \beta {}^G\!\!F_\gamma\partial_\gamma\mathcal{D}_{\alpha\beta}\,.\!\!
\label{eq_D23b1}
\end{align}
Then we introduce the tensors
\begin{align}
\Gamma_{\alpha\gamma}:=&\delta_{\alpha\gamma}-\taua D_\text{t}\beta\,\partial_\gamma {}^G\!\!F_\alpha \label{eq_GammaAPP}\\
\hat{\theta}_{\gamma\beta}:=&\taua D_\text{t} \beta {}^G\!\!F_\delta\partial_\delta\delta_{\gamma\beta} \label{eq_theta}\,
\end{align}
to finally get
\begin{align}
\mathcal{D}_{\gamma\beta}\,\Gamma_{\alpha\gamma}&=\Da G_{\alpha\beta}-\hat{\theta}_{\gamma\beta}\mathcal{D}_{\alpha\gamma}\,.
\label{eq_D23b2}
\end{align}
This differential equation can now be formally solved by means of the alternative iteration scheme \cite{grigolini1986}
\begin{align}
\mathcal{D}^{(n+1)}_{\gamma\beta}\,\Gamma_{\alpha\gamma}&=\Da G_{\alpha\beta}-\hat{\theta}_{\gamma\beta}\mathcal{D}^{(n)}_{\alpha\gamma}\,,
\label{eq_D23b3n}
\end{align}
different from that in Eq.~\eqref{eq_Dn23b}.
Most notably, the 
zero-order term
\begin{align}
\mathcal{D}^{(0)}_{\delta\beta}&=\Da G_{\alpha\beta}\Gamma^{-1}_{\delta\alpha}
\label{eq_D23b3n0}
\end{align}
reduces to the Fox result, Eq.~\eqref{eq_GammaF}, in the absence of a magnetic field and Brownian noise.
Therefore, we have both provided a gradual extension of the multicomponent Fox approach to include higher-order terms
and included the contribution due to an external magnetic field.

\section{The BFPA in two dimensions \label{sec_results2Cnew}}

The predictions of the BFPA for a particle in an external potential in one spatial dimension
have been discussed in Sec.~\ref{sec_1dBFPA}.
For a single particle in a two-dimensional potential landscape, the differential equations of the BFPA, which determine the components of the effective diffusion tensor $\boldsymbol{\mathcal{D}}$,
depend on the symmetry of the system.
In the planar case, the situation is similar to one dimension, i.e., for a given external potential $\phi(x)$ the Eigenvalue $\mathcal{D}_{xx}(x)$ of $\boldsymbol{\mathcal{D}}$, corresponding to the direction of the potential modulation, obeys the same differential equation as in Eq.~\eqref{eq_D23_1d}. The other Eigenvalue $\mathcal{D}_{yy}=\Da$, corresponding to the perpendicular spatial direction $y$,
follows from the constant solution of $\mathcal{D}_{yy}(x)=\Da +\taua D_\text{t}\phi'(x)\mathcal{D}_{yy}'(x)$ and thus turns out to be irrelevant as in the Fox approximation.

Next, we consider the polar case with the external potential $\phi(r)$,
where the Eigenvalues $\mathcal{D}_{rr}$ and $\mathcal{D}_{\varphi\varphi}$ of the effective diffusion tensor $\boldsymbol{\mathcal{D}}$
correspond to the radial $\hat{e}_r$ and polar $\hat{e}_\varphi$ unit vectors,
in contrast to the more general case considered in Sec.~\ref{sec_ELP2d} in presence of a Lorentz force.
Without magnetic field ($\kappa=0$), the general form of the effective potentials in a polar geometry in two dimensions, Eq.~\eqref{eq_phieffB} reduces to 
\begin{equation}
\phi_\text{eff}'=\frac{\phi'}{\mathcal{D}_{rr}}+\frac{\mathcal{D}_{rr}'}{\mathcal{D}_{rr}}+\frac{\mathcal{D}_{rr}-\mathcal{D}_{\varphi\varphi}}{r\,\mathcal{D}_{rr}}\,.
\label{eq_phieffNOB}
\end{equation}
 Within the BFPA, we find the differential equations
 \begin{align}
\!\!\!\!\mathcal{D}_{rr}(r)(1+\taua D_\text{t} \phi''(r))&=\Da +\taua D_\text{t} \phi'(r)\mathcal{D}'_{rr}(r)\,,\!\!\!\!\cr\!\!\!\!
\mathcal{D}_{\varphi\varphi}(r)\left(1+\taua D_\text{t} \frac{\phi'(r)}{r}\right) &=\Da+\taua D_\text{t} \phi'(r)\mathcal{D}'_{\varphi\varphi}(r)\,\!\!\!\!
\label{eq_D23b2MATproject_2d}
\end{align}
 for the two Eigenvalues,
 where the respective expressions in the brackets on the left-hand sides correspond to the Eigenvalues $E_2(r)$ and $E_1(r)$ of the mobility matrix $\boldsymbol{\Gamma}$, compare Eq.~\eqref{eq_EVs}.
 It is thus apparent that the Fox results are recovered at zero order, i.e., when dropping the second terms on the right-hand side.
  Spatial dimensions higher than two can be treated analogously, where additional Eigenvalues of the form $\mathcal{D}_{\varphi\varphi}$ will emerge.
 
 Regarding Eq.~\eqref{eq_D23b2MATproject_2d}, we see that the radial diffusion $\mathcal{D}_{rr}$ 
 is determined by an equation of exactly the same form as in one dimension, Eq.~\eqref{eq_D23_1d}.
 Given the results discussed in Sec.~\ref{sec_1dBFPA}, it is thus not very likely that the solution for $\mathcal{D}_{rr}(r)$ constitutes an improvement over the Fox expression $\mathcal{D}_{rr}(r)=\Da/(1+\taua D_\text{t} \phi'(r)/r)$ the two-dimensional case.
The biggest problem of Eq.~\eqref{eq_D23b2MATproject_2d}, however, stems from the second equation for the polar diffusivity $\mathcal{D}_{\varphi\varphi}$.
Analysis of the Lipschitz condition reveals that the differential equation is mathematically ill defined for practically all interaction potentials of relevance.

\section{Effective Lorentz potentials in two dimensions \label{app_Dcomponents}}

In this appendix we provide the general representations of the effective Lorentz potentials, as introduced in Sec.~\ref{sec_ELP2d}, obtained from expanding the BFPA result.
For a spatially dependent magnetic field in a polar geometry, 
the general effective potential follows from Eq.~\eqref{eq_phieffB}.
Moreover, we take into account the first derivative of $\kappa(r)$ by defining
\begin{align}
K_\kappa(r):=\taua D_\text{t}\phi'(r)\kappa'(r)\kappa(r)\,,
\label{eq_kappar}
\end{align}
such that the generalized ingredients of Eq.~\eqref{eq_phieffB} become
\begin{align}
    \mathcal{D}_{B}&=I_\text{t}+\Da\,\frac{(\kappa^2+E_1)(\kappa^2+1)}{(\kappa^2+E_1E_2)(\kappa^2+1)-(E_1+1)K_\kappa}\,,\!\!\!\!\!\!\cr
    \mathcal{D}_{rr}&=I_\text{t}+\Da\,\frac{E_1(\kappa^2+1)}{(\kappa^2+E_1E_2)(\kappa^2+1)-(E_1+1)K_\kappa}\,,\!\!\!\!\!\!\cr
    \mathcal{D}_{\varphi\varphi}&=I_\text{t}+\Da\,\frac{E_2(\kappa^2+1)-K_\kappa}{(\kappa^2+E_1E_2)(\kappa^2+1)-(E_1+1)K_\kappa}\,,\!\!\!\!\!\!\cr
    \mathcal{D}_{\varphi r}&=\Da\,\frac{(\kappa^2+1)\kappa-\frac{K_\kappa}{\kappa}}{(\kappa^2+E_1E_2)(\kappa^2+1)-(E_1+1)K_\kappa}\,,\!\!\!\!\!\!\cr
    \mathcal{D}_{r\varphi}&=\Da\,\frac{(\kappa^2+1)\kappa}{(\kappa^2+E_1E_2)(\kappa^2+1)-(E_1+1)K_\kappa}\,,\!\!\!\!\!\!
    \label{eq_DcomponentsFULL}
\end{align}
while the Eigenvalues $E_i(r)$ are still given by Eq.~\eqref{eq_EVs}.
It is easily verified that the first three expressions reduce to Eq.~\eqref{eq_Dcomponents} 
and that $\mathcal{D}_{\varphi r}=-\mathcal{D}_{r\varphi}$ if $\kappa'(r)=0$.

Let us now consider a potential $\phi(x)$ and magnetic field $\kappa(x)$ in a planar geometry with the unit vectors
$\hat{e}_x=(1,0)^T$ and $\hat{e}_y=(0,1)^T$.
Then, analogously to the evaluation in Sec.~\ref{sec_ELP2d}, we have $(1+\kappa^2)\,\boldsymbol{G}\cdot \hat{e}_x=\hat{e}_x-\kappa\hat{e}_y$
and $(1+\kappa^2)\,\boldsymbol{G}\cdot \hat{e}_y=\kappa\hat{e}_x+\hat{e}_y$, such that we must consider
$\boldsymbol{\mathcal{D}}=\sum_{i,j}\mathcal{D}_{ij}\hat{e}_i\otimes\hat{e}_j$
 with $i,j\in\{x,y\}$.
The stationary condition $\bvec{j}\cdot\hat{e}_x=0$ for the current
\begin{equation}
 \bvec{j} =- \boldsymbol{G} \cdot \left(\phi'f_N \hat{e}_x+(f_N\boldsymbol{\mathcal{D}}^T)'\cdot\hat{e}_x\right),
\end{equation}
where the dash now denotes the derivative with respect to $x$,
then leads 
to the general representation
\begin{equation}
\phi_\text{eff}'=\frac{\phi'}{\mathcal{D}_{B}}+\frac{\mathcal{D}_{B}'}{\mathcal{D}_{B}}
\label{eq_phieffBplanar}
\end{equation}
for the derivative of the effective potential $\phi_\text{eff}(x)$,
where $\mathcal{D}_{B}:=\mathcal{D}_{xx}+\kappa\mathcal{D}_{xy}$.
This generalized effective diffusivity can be written as
\begin{align}
    \mathcal{D}_{B}&=I_\text{t}+\Da\,\frac{(\kappa^2+1)^2}{(\kappa^2+E_2)(\kappa^2+1)-2K_\kappa}\,,
    \label{eq_DcomponentsplanarFULL}
\end{align}
where the remaining nontrivial Eigenvalue of the mobility matrix, corresponding to $\hat{e}_x$,
reads $E_2(x)=1+\taua D_\text{t} \phi''(x)$ in this quasi-one-dimensional case and $K_\kappa(x)$ has the same form as in Eq.~\eqref{eq_kappar}.
The expression in Eq.~\eqref{eq_DcomponentsplanarFULL} equals the corresponding formula from Eq.~\eqref{eq_DcomponentsFULL} upon setting $E_1=1$ therein and reduces to
\begin{align}
    \mathcal{D}_{B}&=I_\text{t}+\Da\,\frac{\kappa^2+1}{\kappa^2+E_2}\,
    \label{eq_Dcomponentsplanar}
\end{align}
for a spatially constant magnetic field with $\kappa'(x)=0$.

\end{document}